\documentclass[lettersize,journal]{IEEEtran}

\usepackage[colorinlistoftodos]{todonotes}
\usepackage{cite}
\usepackage{amsmath,amssymb,amsfonts}
\usepackage{mathtools}
\usepackage{algorithmic}
\usepackage{float,subfloat}
\usepackage{graphicx}
\usepackage{textcomp}
\usepackage{xcolor}
\usepackage{listings}
\usepackage{amsthm}
\usepackage{changepage}
\usepackage{enumitem}
\usepackage{hyperref}
\usepackage{threeparttable,multicol,makecell,multirow,bm,adjustbox,booktabs,mdframed}
\usepackage{tcolorbox}
\tcbuselibrary{listingsutf8}
\usepackage[normalem]{ulem} 

\newif\ifshowchanges
\showchangesfalse 
\newif\ifshowminorchanges
\showminorchangesfalse
\newcommand{\add}[1]{%
  \ifshowchanges
    {\color{blue}#1}%
  \else
    {#1}%
  \fi
}

\newcommand{\addminor}[1]{%
  \ifshowminorchanges
    {\color{blue}#1}%
  \else
    {#1}%
  \fi
}

\newcommand{\techs}{\textsc{MutGen-S}}
\newcommand{\techf}{\textsc{MutGen-F}}
\newcommand{\techmf}{\textsc{MutGen-MF}}
\newcommand{\tech}{\textsc{MutGen}}
\newcommand{\vanilla}{\textsc{Gen}\textsubscript{vanilla}}
\newcommand{\evo}{\textsc{EvoSuite}}
\newcommand{\evovariant}{\textsc{EvoSuite}\textsubscript{mut}}
\def\BibTeX{{\rm B\kern-.05em{\sc i\kern-.025em b}\kern-.08em
    T\kern-.1667em\lower.7ex\hbox{E}\kern-.125emX}}
\begin{document}

\title{Mutation-Guided Unit Test Generation with a Large Language Model} 

\author{Guancheng Wang\href{https://orcid.org/0000-0002-4338-8813}{\includegraphics[scale=0.06]{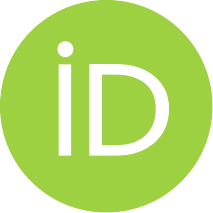}},~\IEEEmembership{Member,~IEEE,}
Qinghua Xu\href{https://orcid.org/0000-0001-8104-1645}{\includegraphics[scale=0.06]{orcid.pdf}},~\IEEEmembership{Member,~IEEE,}
Lionel Briand\href{https://orcid.org/0000-0002-1393-1010}{\includegraphics[scale=0.06]{orcid.pdf}},~\IEEEmembership{Fellow,~IEEE,}
Kui Liu\href{https://orcid.org/0000-0003-0145-615X}{\includegraphics[scale=0.06]{orcid.pdf}},~\IEEEmembership{Member,~IEEE,}
\thanks{Guancheng Wang is with the Research Ireland Lero Centre for Software, University of Limerick, V94 T9PX Limerick, Ireland (e-mail:
guancheng.wang@ul.ie).}
\thanks{Qinghua Xu is with the Research Ireland Lero Centre for Software, University of Limerick, V94 T9PX Limerick, Ireland (e-mail:
qinghua.xu@ul.ie).}
\thanks{Lionel Briand is with the University of Ottawa, Ottawa, ON K1N 6N5,
Canada, and also with the Research Ireland Lero Centre for Software,
University of Limerick, V94 T9PX Limerick, Ireland (e-mail:
lbriand@uottawa.ca, lionel.briand@lero.ie).}
\thanks{Kui Liu is with Software Engineering Application Technology Lab
Huawei, Hangzhou, China (e-mail:
brucekuiliu@gmail.com).}
}

\IEEEpubid{0000--0000/00\$00.00~\copyright~2021 IEEE}

\maketitle

\begin{abstract} 
Unit tests play a vital role in uncovering potential faults in software. While tools like EvoSuite focus on maximizing code coverage, recent advances in large language models (LLMs) have shifted attention toward LLM-based test generation. However, code coverage metrics---such as line and branch coverage---remain overly emphasized in reported research, despite being weak indicators of a test suite’s fault-detection capability. In contrast, \textit{mutation score} offers a more reliable and stringent measure, as demonstrated in our findings where some test suites achieve 100\% coverage but only 4\% mutation score. Although a few studies consider mutation score, the effectiveness of LLMs in killing mutants remains underexplored.

In this paper, we propose \tech{}, a mutation-guided, LLM-based test generation approach that incorporates mutation feedback directly into the prompt. Evaluated on 204 subjects from two benchmarks, \tech{} significantly outperforms both EvoSuite and vanilla prompt-based strategies in terms of mutation score. Furthermore, \tech{} introduces an iterative generation mechanism that pushes the limits of LLMs in killing additional mutants. Our study also provides insights into the limitations of LLM-based generation, analyzing the reasons for live and uncovered mutants, and the impact of different mutation operators on generation effectiveness.
\end{abstract}


\section{Introduction} \label{sec:intro}
\IEEEPARstart{U}{nit} test generation is a fundamental task in software engineering, crucial for ensuring code quality and reducing manual testing effort~\cite{naik2011software,fraser2014large,corno2004automatic,anand2013orchestrated}. With the rapid advancement of Large Language Models (LLMs), research has shifted from traditional techniques—such as random testing~\cite{pacheco2007randoop,hamlet1994random,andrews2011genetic,oriat2005jartege,pacheco2007feedback} and search-based methods~\cite{fraser2011evosuite,rojas2016seeding,rojas2015combining,vivanti2013search}—to LLM-based approaches for generating unit tests~\cite{fan2023large,wang2024software,hou2024large}.

However, most existing work evaluates generated test cases primarily based on code coverage metrics, such as line or branch coverage. According to prior studies~\cite{chekam2017empirical,andrews2006using,inozemtseva2014coverage}, high code coverage does not necessarily imply strong fault detection capability. Other research~\cite{chekam2017empirical,alshahwan2024automated,foster2025mutation} suggests that the mutation score is a more reliable and meaningful metric for evaluating the effectiveness of test cases. Mutation testing introduces artificial faults, known as \emph{mutants}, into the program by applying syntactic changes using mutation operators. The generated test cases are executed against the mutants, and if a test causes a mutant to fail (i.e., ``kills'' it), it is considered effective. The mutation score is the ratio of killed mutants to the total number of generated mutants.

Recent efforts have begun exploring mutation-based evaluation for LLM-generated test cases. For example, Foster et al.~\cite{foster2025mutation} investigate regression testing scenarios in privacy-critical Kotlin applications such as WhatsApp and Instagram. They focus on generating test cases from existing ones and generating mutants using LLMs, which introduces challenges such as detecting equivalent mutants that no test can kill. 
Another work by Dakhel et al.~\cite{dakhel2024effective} targets Python benchmarks and proposes a prompt-based approach to improve mutation score. However, they iteratively add a single mutant until they cannot kill the new mutant. Therefore, they do not attempt to converge towards a maximum mutation score to maximize fault detection. 
Moreover, their study does not analyze why live mutants remain undetected, nor does it consider uncovered mutants, both of which are challenges we address in our research.
\IEEEpubidadjcol

While these efforts represent promising first steps, the use of mutation testing as a core evaluation and improvement strategy for LLM-generated test cases remains underexplored. To address this gap, we propose \tech{}, a mutation-guided LLM-based approach that explicitly maximizes the mutation score. Specifically, a key feature of \tech{} is that we incorporate mutation feedback (i.e., live and uncovered mutant information) into prompt construction to guide the LLM in generating more effective test cases. 
Further, as discussed in prior studies~\cite{tian2025fixing,wang2025projecttest,pan2024multi}, incorporating a fixing step is crucial for improving test generation effectiveness. While prior work primarily focuses on improving metrics such as code coverage or test passing rate, our work explicitly targets the fault detection capability of test cases, measured by the mutation score. To this end, our approach introduces a fixing step that specifically addresses assertion failures in the generated test cases, caused by incorrect use of assertion functions, including erroneous invocations and inappropriate parameter values.

Moreover, we observe that comments in the target code can mislead LLMs into producing irrelevant or incorrect tests. To address this, we introduce a code summarization step during preprocessing that replaces misleading comments with concise summaries, which are then embedded into the prompt. Lastly, since achieving the maximum mutation score is inherently difficult, we introduce an iterative test-generation strategy to push LLMs toward killing more live mutants. Once the mutation score converges, we analyze both uncovered and live mutants, as well as specific mutation operators that remain particularly challenging for the LLM. Like many recent studies~\cite{yuan2024evaluating,gu2025llm,ryan2024code}, our approach focuses on function-level testing. While this prior work also generates test cases for functions from complex datasets such as Defects4J~\cite{just2014defects4j}, it handles cross-module dependencies explicitly by simply providing them in the prompts. In contrast, our work does not aim to address dependency resolution at this stage; instead, it focuses on improving mutation scores for function-level testing. Nevertheless, as discussed below, one of our datasets includes solutions to challenging algorithmic problems with complex program structures, providing a more complex dataset for evaluating the effectiveness of \tech{} in generating test cases.

To evaluate \tech{}, we use two datasets: HumanEval-Java, widely adopted in prior work~\cite{tian2025fixing,dakhel2024effective,siddiq2024using}, and a new LeetCode-Java corpus containing 100 randomly selected algorithmic problems with ground-truth solutions. The experimental results show that test cases generated by \tech{} outperform both \evo{}, a state-of-the-art search-based technique, and a vanilla LLM-based method in terms of mutation score. \tech{} achieves mutation scores of 89.5\% and 89.1\% on HumanEval-Java and LeetCode-Java, respectively, across a total of 1,144 and 1,900 mutants. \addminor{
In addition, although some generated test cases may initially fail, \tech{} is able to repair around 50\% of them, which is comparable to strong results reported in recent software repair literature~\cite{yang2025survey,xia2024automated,gu2024testart}. We further analyze the causes of the remaining failures through a manual inspection of 50 randomly sampled cases, together with the analyses of live and uncovered mutants and the impact of different mutation operators, to better understand the strengths and limitations of LLM-based test generation.}

To understand the contribution of each component in \tech{}, we conduct an ablation study using three variants: \techs{} (without summarization), \techf{} (without the fixing step), and \techmf{} (without mutation feedback). The results confirm the importance of each component in improving the overall mutation score.

In summary, this paper makes the following contributions:
\begin{itemize}
    \item We propose a mutation-guided, LLM-based approach (\tech{}) for automatically generating test cases with high fault detection capability (i.e., mutation score), including an iterative generation process guided by mutation feedback to push the limits of LLMs in killing mutants.
    \item We evaluate our approach on a widely used benchmark dataset and a new dataset we created to increase diversity.
    \item We analyze the effectiveness of \tech{} across different mutation operators and investigate the reasons behind the presence of live and uncovered mutants when \tech{} reaches its maximum mutation score.
\end{itemize}

\section{Methodology}
\label{sec:app}

This section presents our approach \tech{} through a motivating example and a formal description. 
We first illustrate the limitations of standard LLM-based test generation using a running example. We then precisely describe and formalize our methodology, including the complete set of user prompts employed in our approach. To save space and focus on the essential context, we omit the system prompts and the standard formatting rules widely used in existing open-source tools~\cite{mutahunter} and test generation research~\cite{foster2025mutation,chen2024chatunitest,gu2025llm}, which are also available in our open-source implementation.

\subsection{Motivating Example and Intuition}
\label{sec:example_gen}

This subsection goes through a simple example to illustrate the limitations of standard prompting and to motivate our approach.
The example Java code under test shown in Figure~\ref{fig:example-code}, which validates a date format, includes descriptive comments and input-output examples. For clarity and to save space, we retain only the most relevant parts of the comments in the figure. Most LLM-based techniques prompt LLMs using the following standard prompt:
\begin{quote}
\textit{``Generate JUnit tests for the following method to achieve high branch and line coverage.''}
\end{quote}

\begin{figure}[ht]
    \centering
    \begin{lstlisting}[basicstyle=\ttfamily\scriptsize,
  numbers=left,              
  numberstyle=\tiny,
  stepnumber=1,              
  numbersep=4pt,
  frame=single,
  breaklines=true,
  showstringspaces=false,xleftmargin=1.5em,
  tabsize=2,language=Java, label={lst:example-code}]
package original;
class ValidDate {
    /** You have to write a function ......
     * The date is valid if all of the following rules are satisfied:
     * 1-3. ......
     * 4. The date should be in the format: mm-dd-yyyy
     * for example:
     * validDate("03-11-2000") => True
     * validDate("06/04/2020") => False
     * MORE EXAMPLES ...... **/
    public static Boolean validDate(String date) {
        if (date.length() != 10) {
            return false;
        }
        String[] dateArr = date.split("-");
        if (dateArr.length != 3) return false;
        int month = Integer.parseInt(dateArr[0]);
        int day = Integer.parseInt(dateArr[1]);
        int year = Integer.parseInt(dateArr[2]);
        if (month < 1 || month > 12) { return false; }
        if (month == 2) {
            if (day < 1 || day > 29) { return false; }
        } else if (month == 4 || month == 6 || month == 9 || month == 11) {
            if (day < 1 || day > 30) { return false; }
        } else {
            if (day < 1 || day > 31) { return false; }
        }
        return true;
    }
}
\end{lstlisting}
    \caption{Example code under test}
    \label{fig:example-code}
\end{figure}

When combined with the example code (including comments), LLMs can generate test cases that achieve high line and branch coverage. However, as demonstrated in prior work~\cite{foster2025mutation,chekam2017empirical,alshahwan2024automated}, high coverage does not necessarily imply strong fault-detection capability when measured by the mutation score. 
For instance, in our experiments, LLMs generate tests for the subject \texttt{id\_81} from HumanEval-Java with 100\% line and branch coverage, yet the corresponding mutation score is only 4\%.

Listing~\ref{lst:example} shows a representative test function generated by Llama-3.3. In practice, multiple such test functions may be generated, often with a variety of data inputs aimed at maximizing code coverage. However, we observe that most of these inputs fail to cover corner cases, such as February 29th in leap years, likely because such examples are absent from the provided comment block. 
The comments in the target code may also introduce additional issues. For example, the comment begins with “You have to write a function…”, which, when included as part of the prompt, may mislead the LLM into replicating the target code logic rather than generating a meaningful test function. Additionally, Llama-3.3 occasionally produces dates in incorrect formats, such as \texttt{dd-mm-yyyy}, likely due to ambiguity or insufficient specification of the expected format in the program's comments.

Furthermore, when relying on mutation rather than code coverage, the generated inputs often fail to effectively kill mutants. For example, when the \texttt{ConditionalsBoundary} mutation operator in PITest replaces \texttt{<} with \texttt{<=}, it changes the conditional boundary and modifies line 24 as follows:

\begin{lstlisting}[language=Java, basicstyle=\ttfamily\footnotesize, label=lst:mutexample]
if (day <= 1 || day > 30) return false;
\end{lstlisting}
This change introduces a fault: the original code accepts day == 1 as valid, while the mutant incorrectly rejects it. We observe that LLM-generated test cases, when prompted with the original code, tend to focus on representative invalid inputs (e.g., day = 0) and on valid values far from the boundary (e.g., day = 2 or higher). For instance, the model commonly produces a test such as:
\begin{lstlisting}[language=Java, basicstyle=\ttfamily\footnotesize, label=lst:assert_1]
assertFalse(validDate("04-00-2025"));
\end{lstlisting}
which fails in both the original function and the mutant. However, it frequently fails to generate:
\begin{lstlisting}[language=Java, basicstyle=\ttfamily\footnotesize, label=lst:assert_2]
assertTrue(validDate("04-01-2025"));
\end{lstlisting}
which is essential to kill the boundary-mutated version of the code.

\begin{lstlisting}[language=Java, basicstyle=\ttfamily\footnotesize, caption=Example test function generated by Llama-3.3, label=lst:example]
@Test
public void testDayBoundaryInApril() {
  String date = "04-01-2025";
  assertTrue(ValidDate.validDate(date));
}
\end{lstlisting}

In summary, test cases generated using the standard prompt exhibit the following limitations:
\begin{enumerate}
    \item Comments in the target code can mislead LLMs, resulting in the generation of ineffective or irrelevant test cases. 
    \item Given the presence of code mutants, LLMs struggle to understand how to generate tests that kill them effectively. 
\end{enumerate}

To address these issues, we propose \tech{}, an approach that improves the fault detection capability of LLM-generated test cases via a two-stage process: preprocessing and generation, which are detailed in Section~\ref{sec:prep} and Sections~\ref{sec:aug}-\ref{sec:limit}, respectively.

\subsection{Formal Overview of \tech{}}
\label{sec:overview}

\begin{figure*}[htbp]
    \centering
    \includegraphics[width=\textwidth, trim=0cm 0cm 20cm 0cm, clip]{images/overview.pdf}
    \caption{\tech{} Overview}
    \label{fig:overview}
\end{figure*}

Let \( \mathcal{M} = \{m_1, m_2, \dots, m_n\} \) denote the set of mutants generated from the program under test. The mutation score \( \mu(T) \) of a test suite \( T \) is defined as:
\[
\mu(\mathcal{T}) = \frac{|\{ m \in \mathcal{M} \mid \mathcal{T} \text{ kills } m \}|}{|\mathcal{M}|}
\]

We aim to generate a test suite \( \mathcal{T} \) such that \( \mu(\mathcal{T}) \) is maximized. To achieve this, we propose a two-stage process based on prompting an LLM, namely preprocessing and generation.

The preprocessing stage prepares information to guide test case generation, including code summarization and mutation feedback, as detailed in Section~\ref{sec:prep}. In the generation stage, the LLM is prompted with mutation feedback, such as descriptions of mutation operators and diff-like representations of mutants, to generate test cases. 

Let \( \mathcal{R} = \{r_1, r_2, \dots, r_k\} \) denote the set of mutation feedback entries (i.e., killed, live, uncovered), where each \( r_i \) is a tuple \( \langle \text{id}_i, m_i, s_i, \text{op}_i \rangle \), respectively representing the mutant identifier, the mutated statement, the mutant status, and the applied mutation operator. \tech{} first generates an initial test suite \( \mathcal{T} \):
\[
\mathcal{T} = \text{LLM}(\text{Prompt}_{\text{mut}}(\mathcal{R}))
\]

Here, as further described in the following sections, \( \text{Prompt}_{\text{mut}} \) is a structured prompt that encodes details of the mutation operators and specific mutant instances, as illustrated in the example \( \text{Prompt}_{\text{mut}} \) shown in Figure~\ref{fig:prompt-gen}.

After executing \( \mathcal{T} \), \tech{} collects a set \( \mathcal{F}=\{f_1,...,f_q\}  \) consisting of all test cases that encountered execution failures and corresponding error messages. For each $f_i$, \tech{} constructs a second prompt \( \text{Prompt}_{\text{fail}} \) (as illustrated in the example shown in Figure~\ref{fig:prompt-fix}) and invokes the LLM to generate a set of fixed test cases \( \mathcal{T}_{fix} \):
\begin{equation*}
    \mathcal{T}_{fix} = \{ \text{LLM}(\text{Prompt}_{\text{fail}}(f_i)) \mid f_i \in \mathcal{F} \}
\end{equation*}

The final test suite comprises all passing test cases obtained by combining the filtered and fixed tests. We denote this using the function \texttt{ExeFilter}, as follows:
\begin{equation*}
\mathcal{T}_{\text{final}} = \texttt{ExeFilter}((\mathcal{T} \setminus \mathcal{F}) \cup \mathcal{T}_{\text{fix}})
\end{equation*}
This process will be detailed in Section~\ref{sec:fix}.

Our objective is to maximize the mutation score:

\[
\mathcal{T}^* = \arg\max_{\mathcal{T}} \mu(\mathcal{T})
\]

Our assumption, as reported in the mutation testing literature~\cite{chekam2017empirical,andrews2006using}, is that by maximizing the mutation score, we also maximize fault detection effectiveness. In any case, achieving mutation coverage is known to be a more stringent criterion than line or branch coverage~\cite{foster2025mutation} and therefore enables the generation of more effective test suites. However, as revealed in recent work~\cite{gao2025prompt}, finding an optimal prompt for LLMs remains a significant challenge, and we introduce an approximate and iterative solution, detailed in Section~\ref{sec:limit}, which we investigate empirically and whose limitations we analyze, for example, in terms of mutation operators that are difficult to kill.

Figure~\ref{fig:overview} shows an overview of  \tech{}. \tech{} takes a program under test as input and returns test cases aimed at improving its fault-detection effectiveness by maximizing the test suite's mutation score. 
We highlight all data flows in red in the figure.

\subsection{Preprocessing Stage}
\label{sec:prep}
The preprocessing stage prepares information to guide the
test case generation.

\textit{Code Summarization.} 
To address the first limitation above, \tech{} removes comments from the target code and uses a prompt (shown in Figure~\ref{fig:prompt-summary}) to instruct the LLM to summarize the method’s purpose and input format.
The prompt used is as follows:

\begin{figure}[ht]
    \centering
    \resizebox{0.9\linewidth}{!}{
    \begin{tcolorbox}[title=Prompt Used for Summarizing,
colback=gray!5, colframe=black!40!white, fonttitle=\bfseries,
listing only, listing options={
    basicstyle=\ttfamily\small,
    breaklines=true,
    language=Python
}]
Summarize the following Java source code in 1-3 sentences, including its purpose and input format (if applicable).

\centering\texttt{\textbf{\{\{ source code \}\}}}
    \end{tcolorbox}
    }
    \caption{Prompt used for summarizing}
    \label{fig:prompt-summary}
\end{figure}
For the target code shown in Figure~\ref{fig:example-code}, Llama-3.3 returns the following summary:
\begin{quote}
\textit{``This Java function, `validDate', validates a given date string in the format ``mm-dd-yyyy" and returns `true' if the date is valid according to specific rules regarding month and day ranges, and `false' otherwise. The input date string must be non-empty and follow the exact specified format. If any of these conditions are not met, the function will return `false'.''}
\end{quote}
Since our work focuses on improving the fault-detection capability of generated tests through mutation feedback, we directly use the summaries produced by the LLM without further optimization. Investigating how the size and quality of generated summaries affect test generation effectiveness is a promising direction for future work.

\textit{Mutation Feedback.}
To address the second issue, we leverage mutation feedback to guide the LLM in generating test cases with greater fault-detection capability. Before the generation stage, \tech{} extracts mutation feedback from mutation reports produced by mutation testing tools. In this example, we use PITest~\cite{pitest} to generate a mutation report, as shown in Figure~\ref{fig:example-mutreport}. From the report, \tech{} retrieves key information, including the mutation location (line number), the mutant status (killed, survived, or uncovered), and the mutation operator applied to create the mutant. \tech{} then integrates this feedback with the source code, forming a prompt (shown in Figure~\ref{fig:prompt-gen}) that guides the LLM in generating more effective test cases during the generation stage. 

For example, the mutation feedback extracted from the report for the mutant shown in Listing~\ref{lst:mutexample} is as follows:

\begin{quote}
    \textit{21. conditional boundary at line 24, uncovered:}\par
    \texttt{if (day<=1||day>30) return false;}
\end{quote}
This indicates that the existing tests do not cover the 21st mutant, generated by applying a conditional boundary change at line 24, specifically by replacing \texttt{<} with \texttt{<=}, as shown in the second line.
Formally, this mutation feedback entry can be represented as:
\begin{quote}
$r_{21} = \langle 21,$\\
\hspace*{2em} $if\ (day \le 1||day>30)\ return false;$,\\
\hspace*{2em} \textit{uncovered},\\
\hspace*{2em} \textit{change conditional boundary at line 24} $\rangle$
\end{quote}

\begin{figure*}[ht]
\centering
\scriptsize
\begin{adjustwidth}{3cm}{0pt}
\begin{tabular}{@{}l p{0.85\textwidth}@{}}
\multicolumn{2}{l}{\textbf{Mutations}} \\[0.5em]

\colorbox{green!20}{12} & \colorbox{green!20}{1. negated conditional \texttt{->} KILLED} \\
\colorbox{green!20}{13} & \colorbox{green!20}{1. replaced Boolean return with True for original/ValidDate::validDate \texttt{->} KILLED} \\
\colorbox{green!20}{16} & \colorbox{green!20}{1. negated conditional \texttt{->} KILLED} \\
& \colorbox{green!20}{2. replaced Boolean return with True for original/ValidDate::validDate \texttt{->} KILLED} \\[0.5em]

\colorbox{red!20}{20} & \colorbox{red!20}{1. negated conditional \texttt{->} SURVIVED \textcolor{blue}{Covering tests}} \\
& \colorbox{red!20}{2. changed conditional boundary \texttt{->} SURVIVED \textcolor{blue}{Covering tests}} \\
& \colorbox{red!20}{3. changed conditional boundary\texttt{->} SURVIVED \textcolor{blue}{Covering tests}} \\
& \colorbox{red!20}{4. replaced Boolean return with True for original/ValidDate::validDate \texttt{->} SURVIVED \textcolor{blue}{Covering tests}} \\
& \colorbox{red!20}{5. negated conditional\texttt{->} SURVIVED \textcolor{blue}{Covering tests}} \\[0.5em]

\colorbox{red!20}{21} & \colorbox{red!20}{1. negated conditional \texttt{->} SURVIVED \textcolor{blue}{Covering tests}} \\
[0.5em]

\colorbox{red!20}{22} & \colorbox{red!20}{1. negated conditional \texttt{->} SURVIVED \textcolor{blue}{Covering tests}} \\
& \colorbox{red!20}{2. changed conditional boundary \texttt{->} SURVIVED \textcolor{blue}{Covering tests}} \\
& \colorbox{red!20}{3. changed conditional \texttt{->} SURVIVED \textcolor{blue}{Covering tests}} \\
& \colorbox{red!20}{4. replaced Boolean return with True for original/ValidDate::validDate \texttt{->} SURVIVED \textcolor{blue}{Covering tests}} \\
& \colorbox{red!20}{5. changed conditional boundary\texttt{->} SURVIVED \textcolor{blue}{Covering tests}} \\[0.5em]

\colorbox{red!20}{23} & \colorbox{red!20}{1. negated conditional \texttt{->} SURVIVED \textcolor{blue}{Covering tests}} \\
& \colorbox{red!20}{2. negated conditional \texttt{->} SURVIVED \textcolor{blue}{Covering tests}} \\
& \colorbox{red!20}{3. negated conditional \texttt{->} SURVIVED \textcolor{blue}{Covering tests}} \\
& \colorbox{red!20}{4. negated conditional \texttt{->} SURVIVED \textcolor{blue}{Covering tests}} \\[0.5em]

\colorbox{red!20}{24} & \colorbox{red!20}{1. negated conditional \texttt{->} NO\_COVERAGE} \\
& \colorbox{red!20}{2. changed conditional boundary \texttt{->} NO\_COVERAGE} \\
& \colorbox{red!20}{3. changed conditional boundary \texttt{->} NO\_COVERAGE} \\
& \colorbox{red!20}{4. replaced Boolean return with True for original/ValidDate::validDate \texttt{->} NO\_COVERAGE} \\
& \colorbox{red!20}{5. negated conditional \texttt{->} NO\_COVERAGE} \\[0.5em]

\colorbox{red!20}{26} & \colorbox{red!20}{1. changed conditional boundary \texttt{->} NO\_COVERAGE} \\
& \colorbox{red!20}{2. negated conditional \texttt{->} NO\_COVERAGE} \\
& \colorbox{red!20}{3. changed conditional boundary \texttt{->} NO\_COVERAGE} \\
& \colorbox{red!20}{4. replaced Boolean return with True for original/ValidDate::validDate \texttt{->} NO\_COVERAGE} \\
& \colorbox{red!20}{5. negated conditional \texttt{->} NO\_COVERAGE} \\[0.5em]

\colorbox{red!20}{28} & \colorbox{red!20}{1. replaced Boolean return with False for original/ValidDate::validDate \texttt{->} NO\_COVERAGE} \\

\end{tabular}
\end{adjustwidth}
\caption{Example mutation report}
\label{fig:example-mutreport}
\end{figure*}

\subsection{Prompting with Mutation Feedback}
\label{sec:aug}
To maximize the mutation score, \tech{} augments the prompt with mutation feedback collected during the preprocessing stage and uses it to guide Llama-3.3 in generating test cases. 
Figure~\ref{fig:prompt-gen} illustrates the resulting prompt for the HumanEval-Java subject shown in Figure~\ref{fig:example-code}. 
With a single invocation of \tech{}, the generated test suite contains 13 test cases and achieves a mutation score of 70\% over 30 mutants.

We define the prompt construction function as follows:
\begin{align*}
\text{Prompt}_{\text{mut}} & : \mathcal{R} \rightarrow \text{String}, \\
\text{Prompt}_{\text{mut}}(\mathcal{R}) &= \texttt{Format}(desc, src, \bigcup_{r_i \in \mathcal{R}} r_i).
\end{align*}

The function \( \texttt{Format}(\cdot) \) organizes the given parameters into a structured prompt suitable for the LLM.


As illustrated in Figure~\ref{fig:overview} and the example in Section~\ref{sec:example_gen}, the prompt constructed by \tech{} includes three components: a natural language code summarization ($desc$), the source code excluding comments ($src$), and the associated mutation feedback ($\mathcal{R}$). In Figure~\ref{fig:prompt-gen}, the former two correspond to the prompt variables: $desc$ is represented as \textbf{\{\{summary\}\}}, and $src$ as \textbf{\{\{source code without comments\}\}}.

Regarding the content under \textbf{\#\# Mutation Feedback}, we first provide a sentence-level instruction to the LLM. The LLM is provided with live mutants and is prompted to generate tests that are more diverse for each mutant. Following this sentence-level prompt, the set of mutation feedback entries, $\mathcal{R}$, is provided as defined in Section~\ref{sec:prep}.

In addition, \tech{} explicitly specifies the expected output format from the LLM, as illustrated in the prompt example shown in Figure~\ref{fig:prompt-gen}.

\begin{figure}[t!]
  \centering
\resizebox{0.9\linewidth}{!}{\begin{tcolorbox}[title=Prompt Used for Generation,
colback=gray!5, colframe=black!40!white, fonttitle=\bfseries,
listing only, listing options={
    basicstyle=\ttfamily\small,
    breaklines=true,
    language=Python
}]
\# Unit Test Generator for Java using Junit4

\#\# Source Code Summary \texttt{\textbf{\{\{summary\}\}}}

\#\# Source File 

\begin{center}\texttt{\textbf{\{\{source code without comments\}\}}}
\end{center}

\texttt{\textbf{\#\# Mutation Feedback}}

Analyze the live mutant to identify the defects; the generated test cases must look very different from the existing ones.

...

    \textit{21. conditional boundary at line 24, uncovered:}\par
    \texttt{if (day<=1||day>30) return false;}

...

\#\# Output Format in YAML

new\_tests:

\hspace{1em} - test\_behavior: ...

\hspace{3em} test\_name: ...

\hspace{3em} test\_code: ...

\hspace{3em} new\_imports: ...

\end{tcolorbox}
}
  \caption{Prompt used for generation}
  \label{fig:prompt-gen}
\end{figure}

\subsection{Fixing Execution Errors}
\label{sec:fix}

After the initial generation stage, some test cases may fail due to execution errors. \tech{} introduces a \emph{fixing step} that leverages a structured prompt, as illustrated in Figure~\ref{fig:prompt-fix}, to guide the LLM in repairing these failures. In the example shown, two generated test cases initially fail to execute due to runtime errors. By applying the fixing step, \tech{} successfully repairs one of the failing test cases, resulting in a 13\% increase in mutation score.

We categorize execution failures into two types:
\begin{itemize}
    \item \textbf{Assertion Failures:} Test cases that compile successfully but fail at runtime due to violated assertions or incorrect outputs.
    \item \textbf{Compilation Errors:} Test cases that fail to compile due to syntax errors, missing types, unresolved references, or similar issues.
\end{itemize}

To address these failures, \tech{} constructs a structured prompt for each failing test case:
\[
\text{Prompt}_{\text{fail}}(f_i) = \texttt{Format}(t_i, e_i),
\]
where \(t_i\) is the original test case and \(e_i\) is the associated compiler or runtime error message. The prompt explicitly instructs the LLM: 

\emph{"You are processing execution failures. Please match only one of these guidelines and then try to correct the failed tests according to the given error messages."}  

The prompt then enumerates six targeted fix strategies:
\begin{enumerate}
    \item Direct invocation errors: to address incorrect invocation of assertion functions.
    \item \texttt{assertEquals} mismatches: to address oracle value inconsistencies in \texttt{assertEquals} assertions.
    \item \texttt{assertTrue}/\texttt{assertFalse} errors: to address binary oracle inconsistencies in \texttt{assertTrue} or \texttt{assertFalse} assertions.
    \item Ambiguous assertion references: to address ambiguity in \texttt{assertEquals} calls; explicit type casts are applied to clarify the parameter types.
    \item Method name conflicts: to address duplicate method names within the test class.
    \item Other runtime or compilation errors: to address errors not covered by the above strategies.
\end{enumerate}
Detailed instructions for these strategies, along with a fixing example, are provided in Figure~\ref{fig:prompt-fix}.

We define specialized fix functions for each failure type:
\begin{align*}
\text{Fix}_{\text{comp}}(f_i) &= \text{LLM}(\text{Prompt}_{\text{fail}}(f_i)) && \text{for } f_i \in \mathcal{F}_{\text{comp}}, \\
\text{Fix}_{\text{assert}}(f_i) &= \text{LLM}(\text{Prompt}_{\text{fail}}(f_i)) && \text{for } f_i \in \mathcal{F}_{\text{assert}}.
\end{align*}

The outputs of these functions are the repaired test sets:
\begin{align*}
\mathcal{T}' &= \{ \text{Fix}_{\text{comp}}(f_i) \mid f_i \in \mathcal{F}_{\text{comp}} \}, \\
\mathcal{T}'' &= \{ \text{Fix}_{\text{assert}}(f_i) \mid f_i \in \mathcal{F}_{\text{assert}} \},
\end{align*}
and the final set of fixed test cases is:
\[
\mathcal{T}_{\text{fix}} = \mathcal{T}' \cup \mathcal{T}''.
\]

This prompt-based fixing strategy ensures that the LLM systematically addresses different types of execution failures, improving test execution success and contributing to higher mutation scores.

\begin{figure}[ht!]
  \centering
\resizebox{0.9\linewidth}{!}{\begin{tcolorbox}[title=Prompt Used for Fixing,
colback=gray!5, colframe=black!40!white, fonttitle=\bfseries,
listing only, listing options={
    basicstyle=\ttfamily\small,
    breaklines=true,
    language=Python
}]
\#\# Failed Test

The following test failed in the test suite:

\begin{center}
    \texttt{\textbf{\{\{ failed\_test with error message \}\}}}
\end{center}

You are processing execution failures. Please match only one of these guidelines and then try to correct the failed tests according to the given error messages.

	1.	Direct Invocation:
    
	•	Always use assertion functions directly (e.g., assertEquals(value1, value2)) instead of prefixed calls like Assert.assertEquals(value1, value2).
    
	2.	Handling assertEquals Mismatches:
    
	•	If an error follows the pattern:
    
	•	test\_file\_name:line\_number: expected:error\_value but was:correct\_value
    
	•	And involves assertEquals, replace the expected value with the correct one.
    
	•	Example: assertEquals(error\_value, xxx); → assertEquals(correct\_value, xxx);
    
	3.	Fixing assertTrue / assertFalse Errors:
    
	•	If an error follows the pattern:
    
	•	classname:line\_number:null
    
	•	And involves assertTrue or assertFalse, swap the function name while keeping all parameters unchanged.
    
	•	Example: assertTrue(xxx); → assertFalse(xxx);
    
  4. Handling ambiguous 
  
	•	If an error has the pattern:
    
	• reference to assertEquals is ambiguous  
    
	•	And involves assertEquals, add type casts to both parameters based on their respective types.
    
	•	Example: assertEquals(expectedValue, actualValue); → assertEquals((long) expectedValue, (long) actualValue);

  5. Method name conflicts
  
	•	If an error follows the pattern: method is already defined in the class,
    
    •	only append a number to the method name.
    
    •	Example: public void test\_null() → public void test\_null1();

 6. Other errors

	•	Try to fix the errors according to the error messages.

\#\# Output Format in YAML

\textit{The same applied as in Figure~\ref{fig:prompt-gen}.}
\end{tcolorbox}
}
  \caption{Prompt used for fixing}
  \label{fig:prompt-fix}
\end{figure}

\subsection{Pushing the Limit: Iterative Generation}
\label{sec:limit}

We propose an iterative generation process that leverages mutation feedback and failed test cases to progressively enhance the quality and effectiveness of the generated tests. 
At each iteration \( i \), \tech{} obtains a working test suite \( \mathcal{T}_i \).
Before proceeding to the next iteration, \tech{} evaluates \( \mathcal{T}_i \) to obtain mutation feedback \( \mathcal{R}_i\). The LLM is re-invoked with \( \text{Prompt}_{mut}(\mathcal{R}_i)\) to generate an initial set of new test cases. The failing cases are filtered as \( \mathcal{F}_i\) and, for each \(f_j \in \mathcal{F}_i\), the LLM is subsequently invoked with \( \text{Prompt}_{fail}(f_j)\) to fix them. \tech{} obtains \( \Delta \mathcal{T}_i \) as the set of test cases that successfully execute. The updated test suite \( \mathcal{T}_{i+1} \) is defined as:
\[
\mathcal{T}_{i+1} = \mathcal{T}_i \cup \Delta \mathcal{T}_i,
\]

The process continues until convergence (i.e., no significant improvement in mutation score) or until a predefined iteration limit is reached. Once convergence is reached, the remaining live and uncovered mutants provide insights into mutation operators and program constructs that remain challenging for the LLM, which we analyze in Section~\ref{sec:eval_analysis}.

To illustrate the effectiveness of this iterative process, we revisit the running example introduced earlier. After two iterations, for the running example, \tech{} achieves a 100\% mutation score.
In contrast, as demonstrated in Section~\ref{sec:eval}, for this subject in HumanEval-Java, using the vanilla prompt introduced at the beginning of this section, the LLM produces a test suite with a 53\% mutation score, which remains unchanged after four iterations.

\section{Evaluation} \label{sec:eval}
As previously discussed, \tech{} is designed to enhance the fault-detection capability of LLMs in automated test generation by incorporating mutation feedback. To evaluate its effectiveness, we compare \tech{} with three baselines: \textbf{\evo{}}, the state-of-the-art search-based technique; \textbf{\evovariant{}}, a variant that replaces coverage-based fitness functions with strong mutation as described in~\cite{fraser2015achieving}; and \textbf{\vanilla{}}, a prompting-based approach that utilizes an LLM without incorporating mutation feedback and leveraging the comments provided in the program under test through summarization (as illustrated in Figure~\ref{fig:prompt-gen}). We adopt Llama-3.3 70B as the base model in our evaluation due to its recent release, strong effectiveness on code-related tasks~\cite{grattafiori2024llama}, and open-source accessibility, which ensures reproducibility and ease of deployment in both research and practical scenarios.

In addition to effectiveness comparisons, we analyze the limitations of LLM-based test generation by examining killing ratios across different mutation operators. This allows us to identify which mutant types are particularly challenging for LLMs to kill and to offer insights into potential areas for improvement.

Finally, we conduct an ablation study to evaluate the contribution of each component within \tech{} to the overall effectiveness. 
 
In sum, our evaluation addresses the following research questions.
\begin{itemize}
    \item \textbf{RQ1:} How does \tech{} perform in generating test cases with high fault detection capability (i.e., mutation score) compared to \evo{}, \evovariant{}, and \vanilla{}?
    \item \textbf{RQ2:} Which mutation operators remain challenging for LLM-based test generation and for what reasons?
    \item \textbf{RQ3:} What is the individual contribution of each component of \tech{} to the overall effectiveness?
\end{itemize}

\subsection{Experiment Setup} \label{sec:eval_setup}

\textbf{Subjects.} In our study, we primarily focus on method-level subjects, where each subject is a single Java class containing only one target method. We generate unit test cases specifically for this method. We use a widely adopted benchmark dataset that has been used in prior publications~\cite{siddiq2024using,zheng2023codegeex,tian2025fixing,islam2405mapcoder}, and we also create an additional dataset to enable a more diverse evaluation. 
More specifically, we include the following subjects:
\setcounter{paragraph}{0}
\paragraph{HumanEval-Java} Following prior work~\cite{islam2405mapcoder,siddiq2024using,tian2025fixing,zheng2023codegeex}, we adopt the HumanEval dataset~\cite{chen2021evaluating,athiwaratkun2022multi} as one of our evaluation benchmarks. Specifically, we utilize all 160 Java subjects from the dataset. For each subject, we apply \evo{} with its default configuration to automatically generate a test suite. We then evaluate the mutation score of the generated test suite using PITest~\cite{pitest}. 
Subjects where \tech{} and \evo{} both achieved a 100\% mutation score are discarded, resulting in 104 Java subjects for our evaluation. Indeed, these subjects do not help us distinguish the two approaches.
\paragraph{Leetcode-Java} We search GitHub using the keyword \texttt{"Leetcode Java"} and select the top two repositories based on the \emph{Best match} sorting criterion. These repositories contain Java solutions to algorithmic problems from LeetCode~\cite{leetcode}, which are categorized into three difficulty levels: \emph{Easy}, \emph{Medium}, and \emph{Hard}. We exclude all solutions corresponding to Easy problems. 
We collect all Java source files for the remaining \emph{Medium} and \emph{Hard} problems and organize them into a unified package structure following the convention used in HumanEval-Java. We use the Maven build system to compile the project and remove any subjects that failed to compile. For each valid subject, we apply \evo{} with its default settings to generate a test suite. Subjects for which the generated test suite achieved a 100\% mutation score by \tech{} and \evo{} are once again discarded. Finally, for computational feasibility, we randomly select 50 subjects from each of the \emph{Medium} and \emph{Hard} sets, resulting in a total of 100 Leetcode-Java subjects used in our evaluation.

The average number of lines of code in HumanEval-Java methods is 41, with an average cyclomatic complexity (CC) of 4.90, which is close to the average CC of 5.46 reported for typical Java methods~\cite{cc}. This indicates that HumanEval-Java exhibits representative control-flow complexity for standard Java methods. In contrast, methods in Leetcode-Java are generally more complex, with an average CC of 7.88. These statistics demonstrate that Leetcode-Java contains methods with complexity levels comparable to or exceeding those found in real-world Java code~\cite{ccpaper}, making it both representative and more challenging than HumanEval-Java.

\textbf{Metrics.} 
Since our work focuses on enhancing the effectiveness of LLM-generated test cases, we adopt the \textbf{mutation score}, calculated as the ratio of killed mutants to the total number of mutants, as the primary evaluation metric (see Equation~\ref{eq:mutation_score}).

\begin{equation}
\text{Mutation Score} = \frac{\text{\# killed mutants}}{\text{\# total mutants}} \times 100\%
\label{eq:mutation_score}
\end{equation}

A mutant is considered \emph{killed} if at least one generated test case causes a behavioral difference between the original program and its mutated version, typically resulting in a failed assertion or exception.
In addition to the mutation score, we also report \textbf{branch coverage} and \textbf{line coverage}, in line with previous work~\cite{alagarsamy2024a3test,chen2024chatunitest}.

\textbf{Process.} 
To address \textbf{RQ1}, we use all subjects from both datasets. We begin by generating test cases with both \vanilla{} and \tech{} through a single end-to-end execution, recording the execution time and the number of tokens consumed. We then apply \evo{} and \evovariant{} to generate test cases for each subject, setting their time budget to the larger average processing times of \tech{} on the two datasets.
It is important to note that, unlike prior work~\cite{foster2025mutation}, our approach generates tests entirely from scratch, without relying on any existing test cases.
We evaluate each test suite using two structural coverage metrics, \textbf{branch coverage} and \textbf{line coverage}, measured by the JaCoCo tool~\cite{jacoco}. Additionally, we use PITest~\cite{pitest} with its DEFAULTS mutation operator group to measure the \textbf{mutation score} of each test suite. 
The DEFAULTS group provides a set of operators commonly used in the literature~\cite{fraser2015achieving,wang2024comprehensive} and, according to the PITest official documentation~\cite{pitest}, are not easy to detect and minimize the number of equivalent mutants.
While PITest also offers a STRONGER operator group, it differs from DEFAULTS by only two additional mutation operators, whose applicability is rather limited. For example, the \textit{Remove Conditionals} operator only targets equality checks. For the ALL group provided by PITest, the additional mutation operators generally generate mutants that are relatively easy to kill, making them less informative for our evaluation. Note that although PITest is designed for Java, its mutation operators can be applied to other languages as well~\cite{pitest,wang2024comprehensive,jia2010analysis}.
To assess the statistical significance of the improvements achieved by our approach in terms of effectiveness, we performed non-parametric hypothesis testing using Vargha and Delaney’s A12 statistic~\cite{arcuri2011practical,vargha2000critique}. The A12 effect size was computed from the mutation score, which ranges from 0 to 1: a value of 0.5 indicates no difference (i.e., results are due to chance), whereas a value greater than 0.5 indicates that \tech{} is more likely to outperform the compared approach. Conversely, a value less than 0.5 indicates the opposite. We also conducted a paired-sample Wilcoxon signed-rank test on both coverage and mutation score to evaluate whether \tech{} achieves statistically significant improvements on these metrics.

To address \textbf{RQ2}, we analyze the mutation operators involved in the evaluation. 
As a preliminary experiment, we first randomly select 10 subjects from each dataset and perform seven rounds of end-to-end test generation. In each round, \tech{} generates additional test cases from previously generated ones. By tracking mutation scores across iterations, we observe that they tend to converge after the fourth iteration. Therefore, we set the maximum number of repetitions to four in our experiments. We collect results at the fourth iteration, and analyze the distribution of mutation operators across all generated mutants and compute the \textit{killing ratio} for each operator, that is, the proportion of killed mutants relative to the total number of mutants produced by that operator. This allows us to identify which mutation operators tend to produce mutants that are difficult or easy for LLM-generated test cases to kill.
In our evaluation, mutation testing is conducted using PITest, which employs 11 mutation operators as described in~\cite{mutops}.
For mutation operators with low killing ratios, we investigate why LLMs struggle to kill these mutants.
Due to space constraints, we do not present a detailed analysis for each mutation operator or for all live and uncovered mutants. Instead, we analyze a subset of cases and identify several factors that contribute to LLMs' inability to kill certain mutants. Nevertheless, our findings offer valuable insights and highlight important directions for future research in this area.

To address \textbf{RQ3}, we conduct an ablation study to evaluate the contribution of each core component in \tech{}. Specifically, \tech{} consists of three main components: (1) extracting and incorporating code summarization into prompts, (2) applying the fixing step to repair invalid test cases, and (3) extracting and incorporating mutation feedback into prompts. In this study, we disable one component at a time, yielding three variants: \techs{} (using the original code comments instead of summarization), \techf{} (omitting the fixing step), and \techmf{} (excluding mutation feedback). For each variant, we perform four iterations of test case generation and measure mutation score to assess the impact of each component on the effectiveness of \tech{}.

\textbf{Implementation.} We describe important implementation details of \tech{} and our evaluation setup. 
All subjects used in the evaluation are built using the Maven build system. 
We follow official guidelines to integrate \evo{}, JaCoCo, and PITest into the Maven pipeline for automated test generation, code coverage measurement, and mutation testing, respectively. 
Our approach, \tech{}, is implemented in Python and built on top of the Mutahunter framework~\cite{mutahunter}. Specifically, we reuse components from Mutahunter for prompt handling, LLM interfacing, and response processing.
We use Ollama~\cite{ollama} to locally deploy the Llama-3.3 70B model~\cite{grattafiori2024llama} with the default configuration, setting the temperature to 0.0 for more deterministic outputs.
As our main baseline, \evo{} officially and stably supports JUnit 4, while its support for JUnit 5 is currently limited and experimental. Therefore, JUnit 4 remains the recommended choice for reliable test generation. To ensure a fair comparison, we adopt JUnit 4 in all experimental settings.

The evaluation is conducted on a Linux server equipped with a 56-core CPU, 125 GB of RAM, a single NVIDIA RTX 6000 Ada GPU, and running Ubuntu Linux 24.04.1 LTS. 
Our tool and benchmarks can be found at: \href{https://github.com/Amocy-Wang/MUTGEN}{https://github.com/Amocy-Wang/MUTGEN}.

\subsection{Results and Analysis} \label{sec:eval_analysis}
\begin{table*}[t!]
\centering
\caption{Summary of Results: Comparison of \tech{} and baseline approaches on different datasets. For each approach and dataset, we report the average per-subject line coverage (Line Cov), branch coverage (Branch Cov), mutation score (Mutation Score), number of mutants per subject (\#Mutants per Subject), Vargha and Delaney’s A12 effect size (A12 effect size), total runtime in seconds (Time (secs)), and the total number of tokens consumed (\#Tokens). ``*'' denotes cases when \tech{} achieves significantly better results than the compared baselines.}
\vspace{-1em}
\begin{tabular}{ccccccccc}
\hline
Approach & Dataset & Line Cov & Branch Cov & Mutation Score & \#Mutants per Subject & A12 effect size & Time (secs) & \#Tokens\\
\hline
\multirow{2}{*}{\evo{}} 
  & HumanEval-Java & 95.6\% & 93.4\% & 69.5\%    & 11 & 0.759 & 150 & 1240\\
  & Leetcode-Java & \textbf{99.0\%}  & \textbf{98.9\%}     & 58.9\%  & 19  & 0.899 & 150 & 956\\
\hline
\multirow{2}{*}{\evovariant{}} 
  & HumanEval-Java & 95.4\% & 93.4\% & 67.4\%    & 11 &  0.744 & 150 & 1232\\
  & Leetcode-Java & 98.1\%  & 98.7\% & 58.1\%  & 19  &  0.889 & 150 & 940\\
\hline
\multirow{2}{*}{\vanilla{}} 
  & HumanEval-Java  & 96.2\%     & 92.8\%     & 77.9\%  & 11   & 0.650 & 74.1 & 846\\
  & Leetcode-Java & 96.3\%    & 92.7\%    & 69.9\%  & 19  & 0.734 & 76.3 & 1232\\
\hline
\multirow{2}{*}{\tech{}} 
  & HumanEval-Java  & \textbf{98.3\%}     & \textbf{95.8\%}     & \textbf{89.5\%*}  & 11 & -- & 125.9 & 899\\
  & Leetcode-Java &  98.4\%    & 94.8\%   & \textbf{89.1\%*} & 19 &  -- & 149.4 & 1629\\
\hline
\end{tabular}
\label{table:summary}
\end{table*}

\paragraph{Comparison with \evo{}, \evovariant{}, and \vanilla{}}
Table~\ref{table:summary} presents the overall results of \evo{}, \evovariant{}, \vanilla{}, and our approach \tech{} on the two datasets: HumanEval-Java and Leetcode-Java. The best results for each metric and dataset are highlighted.

On HumanEval-Java, for both \textbf{line coverage} and \textbf{branch coverage}, \tech{} outperforms \evo{}, \evovariant{} and \vanilla{}.  
On Leetcode-Java, we see the opposite: \evo{} achieves the highest coverage scores, surpassing both LLM-based techniques.  For both datasets, however, differences are relatively small, with values below 3 percentage points. 
Although \tech{} is primarily designed to improve mutation score, the results also demonstrate that it consistently outperforms \vanilla{} in terms of line and branch coverage.
This supports previous observations ~\cite{foster2025mutation} that improvements in mutation score can also lead to enhanced code coverage.

Regarding the \textbf{mutation score}, PITest generates an average of 11 and 19 mutants per subject on HumanEval-Java and Leetcode-Java, respectively, resulting in a total of 1,144 and 1,900 mutants across the two datasets. \tech{} achieves mutation scores of 89.5\% and 89.1\% on the two datasets. Compared to \evo{}, this represents large improvements of 28.8\% and 51.3\%, respectively, corresponding to around 20 and 30 percentage points. Compared to \vanilla{}, \tech{} also significantly improves the mutation score, though to a lesser extent.
Additionally, \tech{} also surpasses \evovariant{}, which specifically aims to maximize the mutation score, to an even larger extent. Indeed, surprisingly, we observe that \evo{} slightly outperforms \evovariant{}. This might be due to the latter being developed 10 years ago~\cite{fraser2014large} and receiving no further maintenance. For example, we found that the mutation operators used by \evovariant{} are limited, resulting in poor performance of the generated test cases when evaluated by PITest. Another tentative explanation is that, to ensure fair comparisons, recall that we set the time budget to the maximum used by \tech{}, but a larger budget may be necessary for \evovariant{}.

Furthermore, we report the \textit{A12 effect size}~\cite{arcuri2011practical} on mutation scores in the third-to-last column of Table~\ref{table:summary}. For the comparison between \evo{} and \tech{}, the A12 values are 0.759 and 0.899, indicating that \tech{} has a 75.9\% and 89.9\% probability of achieving higher mutation scores than \evo{} on HumanEval-Java and Leetcode-Java, respectively. Similarly, compared with \vanilla{}, the A12 values are 0.650 and 0.734, indicating that \tech{} outperforms \vanilla{} with probabilities of 65.0\% and 73.4\% on the same datasets. Additionally, based on the results of the paired-sample Wilcoxon signed-rank test, we found that \tech{} achieves statistically significant improvements ($\alpha$ = 0.05) in mutation scores compared with \evo{}, \evovariant{}, and \vanilla{}. However, there is no statistically significant difference in coverage, as most subjects already achieve near 100\% coverage across all approaches.

In terms of overhead, \tech{} consumes fewer tokens on HumanEval-Java but more tokens on Leetcode-Java. The LeetCode-Java subjects consist of solutions to medium- and hard-level algorithmic problems, with more complex logic than those in HumanEval-Java. As a result, PITest generates more mutants for them (19 vs. 11), and Llama-3.3 requires more test cases to kill all mutants.
Compared with \vanilla{}, \evo{}, and \evovariant{}, in Leetcode-Java, \tech{} consumes more tokens while achieving higher mutation scores as well as better line and branch coverage. In contrast, on HumanEval-Java, \tech{} consumes fewer tokens than \evo{} and \evovariant{} while still achieving superior performance in both coverage and mutation score. For the baseline approaches (i.e., \evo{} and \evovariant{}), a possible explanation for their higher token consumption on HumanEval-Java is that these programs contain more lines of code on average, whereas LeetCode-Java programs exhibit higher cyclomatic complexity. Longer programs tend to yield larger, more verbose test suites generated by these search-based techniques, resulting in increased token usage at the test level. Interestingly, we observe an opposite trend for our LLM-based approach. On LeetCode-Java, \tech{} consumes more tokens than on HumanEval-Java, despite the latter having longer programs on average. We hypothesize that this behavior is related to the use of mutation feedback in the prompt. Specifically, LeetCode-Java has more mutants per subject, resulting in more uncovered or live mutants included in the mutation feedback provided to the LLM. As a consequence, the prompt becomes richer and more informative, encouraging the LLM to generate more extensive test cases to kill these mutants, which, in turn, leads to higher token consumption.  However, differences of that scale in the number of tokens do not have practical implications. 

Overall, the results suggest that \tech{} significantly improves mutation scores relative to alternatives, with large differences, at the expense of slightly lower code coverage than \evo{} for the most complex subject (Leetcode-Java). 
This is expected, as \evo{} is designed to maximize coverage. For the other subject (HumanEval-Java), the reverse trend is observed. 
Even when \evo{} is specifically configured as \evovariant{} to maximize mutation scores~\cite{fraser2014large}, it performs slightly worse than \evo{} on both datasets.
Therefore, to optimize mutation scores and, thus, fault detection, we recommend using \tech{}, though the impact on coverage is unclear and likely negligible in practice.

\tcbset{
  myRQStyle/.style={
    colback=gray!10,  
    colframe=gray!10, 
    fonttitle=\bfseries,
    coltitle=black,
    boxrule=0.8pt,
    arc=3mm,
    left=2mm,
    right=2mm,
    top=1mm,
    bottom=1mm,
  }
}

\begin{tcolorbox}[myRQStyle, title=Answer to RQ1]
\tech{} consistently achieves higher mutation scores than the baselines, demonstrating its superior fault detection capability. While this may result in a slight decrease in code coverage for more complex subjects (e.g., Leetcode-Java), the difference is practically negligible.
\end{tcolorbox}

\begin{table}[ht]
\centering
\caption{Mutant Status Counts per Operator on HumanEval-Java (Sorted by Total)}
\begin{tabular}{lccccc}
\toprule
\textbf{Operator} & \textbf{K} & \textbf{L} & \textbf{NoCOV} & \textbf{Total} & \textbf{Killing ratio} \\
\midrule
NegateConditionals & 398 & 7 & 3 & 408 & 97.5\% \\
Math & 209 & 11 & 3 & 223 & 93.7\% \\
ConditionalsBoundary & 186 & 18 & 4 & 208 & 89.4\% \\
EmptyReturns & 71 & 1 & 4 & 76 & 93.4\% \\
TrueReturns & 45 & 7 & 2 & 54 & 83.3\% \\
Increments & 50 & 2 & 1 & 53 & 94.3\% \\
PrimitiveReturns & 29 & 4 & 1 & 34 & 85.3\% \\
FalseReturns & 26 & 0 & 1 & 27 & 96.3\% \\
VoidMethodCalls & 18 & 7 & 1 & 26 & 69.2\% \\
NullReturns & 19 & 0 & 0 & 19 & 100.0\% \\
InvertNegatives & 16 & 0 & 0 & 16 & 100.0\% \\
\midrule
Summary & 1067 & 57 & 20 & 1144 & 93.3\% \\
\bottomrule
\end{tabular}
\label{tab:mutant_stats_sorted_humaneval}
\end{table}

\begin{table}[ht]
\centering
\caption{Mutant Status Counts per Operator on Leetcode-Java (Sorted by Total)}
\begin{tabular}{lccccc}
\toprule
\textbf{Operator} & \textbf{K} & \textbf{L} & \textbf{NoCOV} & \textbf{Total} & \textbf{Killing ratio} \\
\midrule
Math & 766 & 8 & 14 & 788 & 97.2\% \\
NegateConditionals & 474 & 5 & 14 & 493 & 96.1\% \\
ConditionalsBoundary & 288 & 11 & 1 & 301 & 95.7\% \\
PrimitiveReturns & 97 & 14 & 2 & 113 & 85.8\% \\
VoidMethodCalls & 58 & 2 & 7 & 67 & 86.6\% \\
Increments & 46 & 7 & 0 & 53 & 86.8\% \\
EmptyReturns & 24 & 0 & 2 & 26 & 92.3\% \\
TrueReturns & 15 & 0 & 11 & 26 & 57.7\% \\
NullReturns & 17 & 0 & 2 & 19 & 89.5\% \\
FalseReturns & 13 & 0 & 0 & 13 & 100.0\% \\
InvertNegatives & 1 & 0 & 1 & 2 & 50.0\% \\
\midrule
Summary & 1799 & 47 & 54 & 1900 & 94.7\% \\
\bottomrule
\end{tabular}
\label{tab:mutant_stats_sorted_leetcode}
\end{table}

\paragraph{Effectiveness of \tech{} on Different Mutation Operators}
To evaluate the effectiveness and limitations of \tech{} in killing mutants, we analyze the killing ratio across different mutation operators (described in~\cite{mutops}) after running \tech{} for four iterations. Tables~\ref{tab:mutant_stats_sorted_humaneval} and \ref{tab:mutant_stats_sorted_leetcode} present the mutant status counts for each mutation operator on HumanEval-Java and Leetcode-Java, respectively. In both tables, Column \textbf{K} indicates the number of killed mutants for each corresponding mutation operator. Column \textbf{L} shows the number of live mutants (covered but not killed), while Column \textbf{NoCOV} represents the number of mutants that are not covered by any test cases. Column \textbf{killing ratio} denotes the proportion of killed mutants relative to the total number of mutants.

\addminor{\tech{} achieves overall killing ratios of 93.3\% and 94.7\% on HumanEval-Java and Leetcode-Java, respectively. \tech{} demonstrates consistently high effectiveness across most mutation operators. For example, after four iterations, the killing ratio for \texttt{Math} reaches 97.2\% on LeetCode-Java; the remaining mutants are predominantly uncovered, indicating that they are not exercised by the generated test cases.}


\add{In contrast, several other mutation operators, including \texttt{VoidMethodCalls} and \texttt{TrueReturns}, exhibit lower effectiveness. For example, \texttt{VoidMethodCalls} shows a killing ratio of 69.2\% on HumanEval-Java, and \texttt{TrueReturns} achieves only 57.7\% on Leetcode-Java.}
\add{Notably, although these operators differ in their occurrence frequency, our correlation analysis indicates no strong and consistent monotonic relationship between operator frequency and killing ratio. The Spearman correlation is $-0.214$ on HumanEval-Java and $0.36$ on LeetCode-Java. These results suggest that frequency alone does not explain the observed differences in effectiveness and that other factors, such as oracle difficulty or semantic complexity, may play a more important role.}

\begin{figure}[t]
  \centering
  \begin{minipage}[b]{0.45\textwidth}
    \centering
    \includegraphics[width=0.75\textwidth]{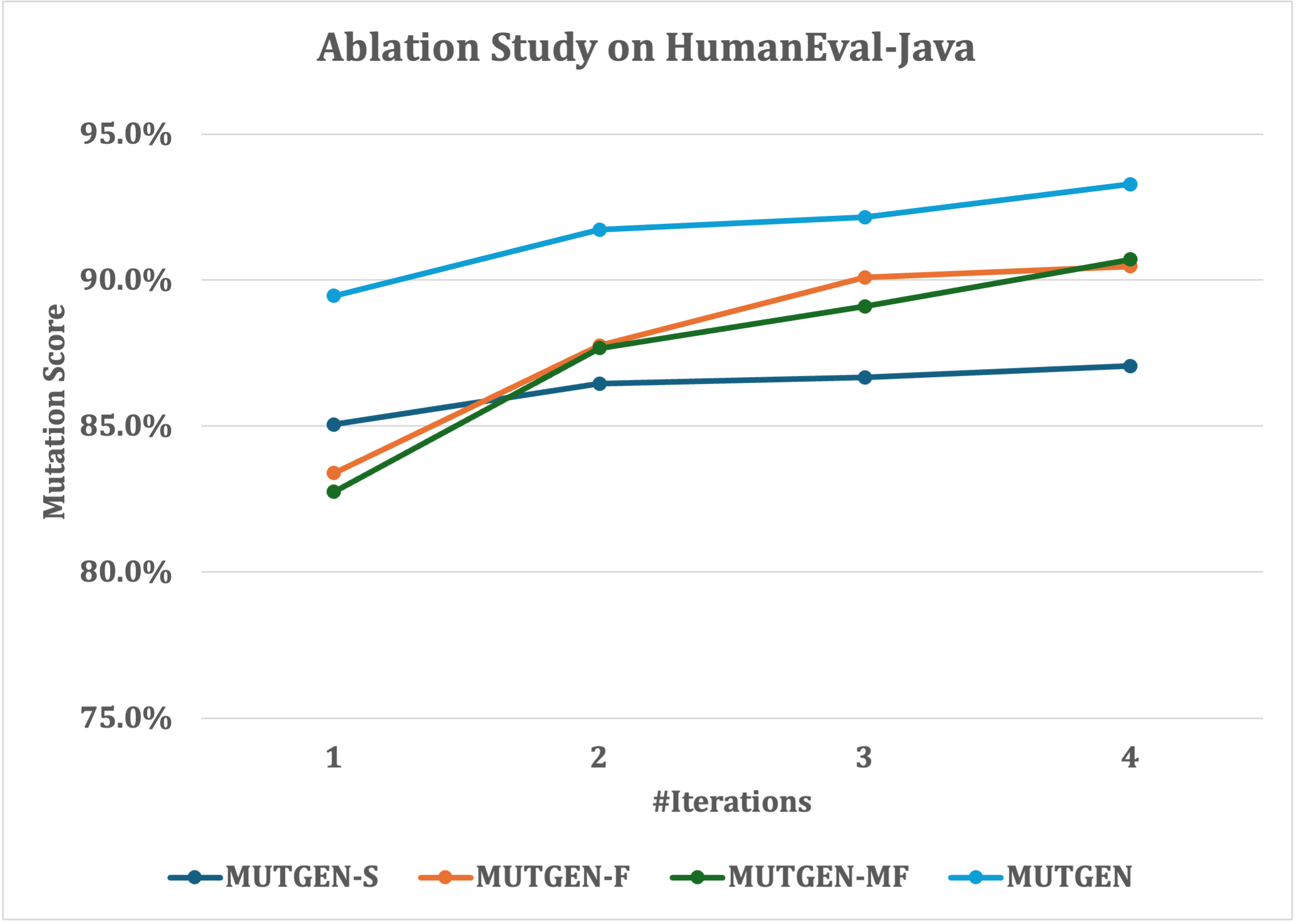}
    \par\vspace{0.5ex}
    {\small (a) Results on HumanEval-Java}
    \label{fig:results_humaneval}
  \end{minipage}
  \hfill
  \begin{minipage}[b]{0.45\textwidth}
    \centering
    \includegraphics[width=0.75\textwidth]{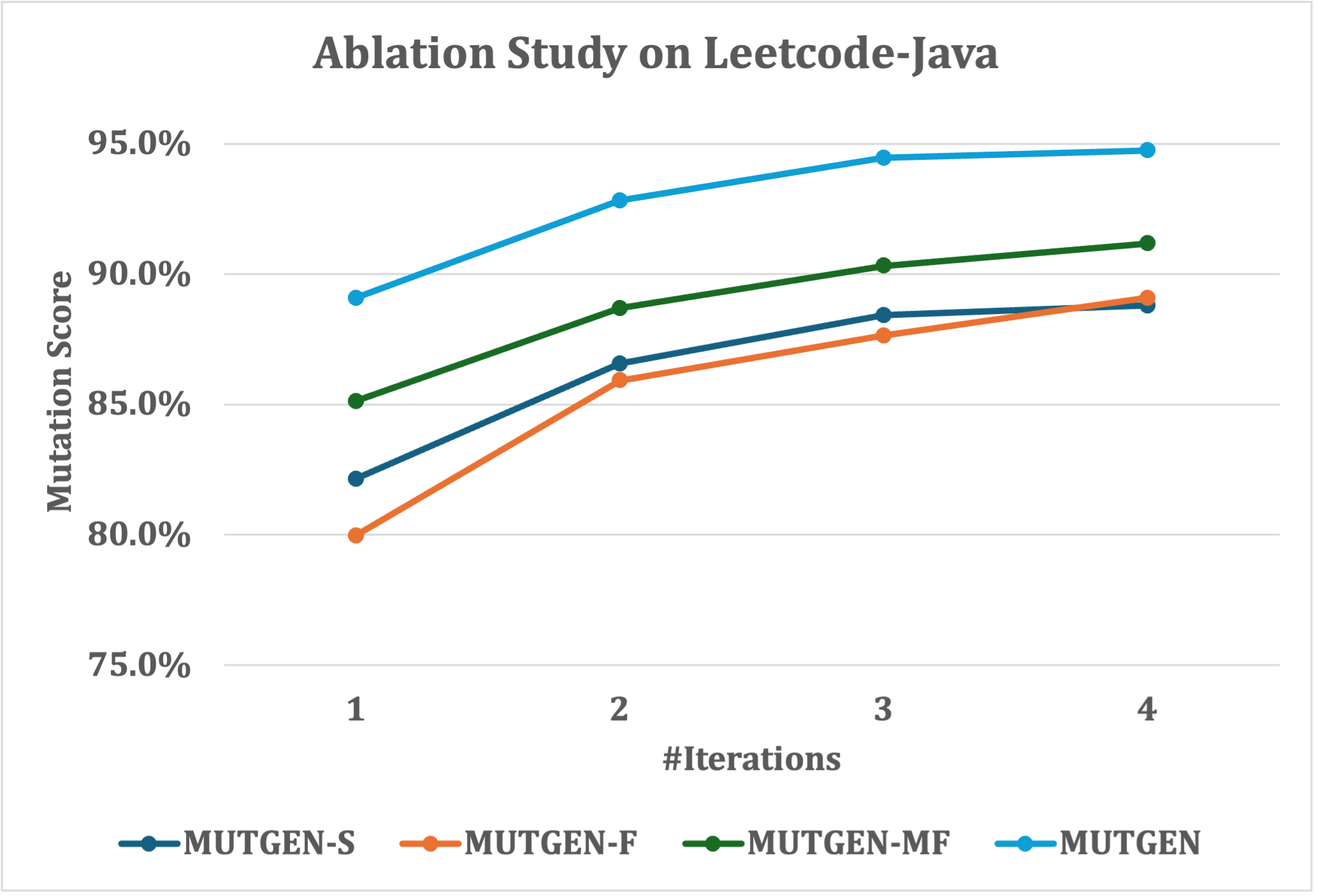}
    \par\vspace{0.5ex}
    {\small (b) Results on Leetcode-Java}
    \label{fig:results_leetcode}
  \end{minipage}
  \caption{Mutation Score Changes over 4 Iterations: Ablation Study on Both Datasets.}
  \label{fig:results_ablation}
\end{figure}

\addminor{
We further analyze the reasons behind live and uncovered mutants, with a particular focus on why certain operators exhibit relatively low killing ratios.

For live mutants, most mutation operators alter the original program logic. For example, the \texttt{MATH} operator may mutate a statement \texttt{a = b + c;} to \texttt{a = b - c;}, thereby changing the value of variable \texttt{a}. If \texttt{a} is not critical to the program’s behavior, the LLM may fail to recognize the impact of the mutation and thus generate test cases that do not expose the fault. This observation underscores the need for finer-grained semantic information to guide LLMs in identifying and targeting meaningful behavioral changes, thereby improving their ability to detect live mutants. 

Similarly, on HumanEval-Java, the \texttt{VoidMethodCalls} operator achieves a killing ratio of 69.2\%, producing 26 mutants, of which 7 remain live. This operator removes calls to void methods, which may alter program behavior in subtle ways that are difficult for the LLM to capture, especially when the side effects of these calls are not explicitly reflected in the output. Addressing such cases requires better modeling of implicit side effects, which may further improve the killing ratio.

In addition to live mutants, a portion of mutants remain uncovered (denoted as \textbf{NoCOV} in our tables), meaning they are not exercised by any generated test cases. Note that these uncovered mutants are distinct from live mutants. The primary cause is dead code and the inability of generated test cases to reach mutated statements, even after incorporating mutation feedback. This suggests that further improvements in mutation score require more fine-grained information and targeted guidance.
}

\begin{tcolorbox}[myRQStyle,   
                  title=Answer to RQ2]
\tech{} achieves overall high killing ratios (93.3\% on HumanEval-Java and 94.7\% on Leetcode-Java), indicating strong effectiveness across most mutation operators. While the majority of operators achieve very high killing ratios, a few, such as \texttt{VoidMethodCalls} and \texttt{TrueReturns}, remain challenging. These exceptions are primarily due to dead code, uncovered statements, or subtle semantic changes that are difficult for the LLM to capture, highlighting the need for more fine-grained semantic guidance.
\end{tcolorbox}


\paragraph{Ablation Study} Figure~\ref{fig:results_ablation} presents the mutation scores achieved by the ablation variants---\techs{}, \techf{}, \techmf{}---and our approach \tech{} across four iterations. Specifically, Figure~\ref{fig:results_ablation}(a) shows the results on HumanEval-Java, while Figure~\ref{fig:results_ablation}(b) reports the results on Leetcode-Java. The horizontal axis represents the number of iterations, and the vertical axis denotes the mutation score.

Based on the mutation scores obtained at the fourth iteration, code summarization appears to have the most significant impact on the effectiveness of \tech{}. As discussed in Section~\ref{sec:example_gen}, code comments can sometimes mislead LLMs or omit essential details—such as input formats—that are crucial for guiding test case generation. 

\add{The fixing stage has a notable impact on effectiveness, particularly on the LeetCode-Java dataset, where many generated test cases initially fail to execute. Prior work reports that LLM-generated tests often suffer from low execution success rates. Our results provide quantitative evidence of improvement from fixing, based on the failed test cases generated across all iterations of MUTGEN.
On HumanEval-Java, MUTGEN produces 1,254 failed test cases, of which 52.08\% are successfully repaired, leaving 601 unfixed test cases. On LeetCode-Java, 47.29\% of the 2,742 failed test cases are fixed, leaving 1,445 unfixed tests. These repaired tests substantially improve mutation scores. \addminor{Note that a repair rate of around 50\% is considered a strong result in the recent software repair literature~\cite{gu2024testart,yang2025survey,xia2024automated}.}
We further manually inspected 50 randomly selected unfixed test cases from the remaining failures. We found that 74\% are caused by incorrect oracles (e.g., wrong assertion functions or erroneous expected outputs), while 26\% are due to runtime exceptions triggered by invalid inputs (e.g., null pointer parameters). This suggests that most remaining failures stem from oracle-level semantic issues, which warrants further work on more advanced fixing strategies.}

Lastly, mutation feedback also enhances effectiveness. As revealed in RQ2, although many mutants generated by common mutation operators are relatively easy for LLMs to kill, mutation feedback still yields significant gains in effectiveness by guiding the model. Future work may further investigate how mutation feedback can be tailored to help LLMs kill harder-to-detect mutants and which mutation operators benefit most from such guidance. We leave this exploration as a direction for future research.

\begin{tcolorbox}[myRQStyle, title=Answer to RQ3]
Among MUTGEN’s components, code summarization has the greatest impact on effectiveness by providing clear guidance to the LLM; the fixing stage increases the number of test cases that execute successfully, thereby improving the mutation score; and mutation feedback further improves performance by guiding the LLM to target harder-to-kill mutants.
\end{tcolorbox}

\subsection{Threats to Validity} \label{sec:eval_threats}
The primary threat to internal validity lies in the correctness of the implementation of \tech{} and the experimental scripts. To reduce this threat, we have carefully reviewed our code. Another potential threat to internal validity concerns the configuration of the employed LLM, as variations in parameters such as temperature, top-p, and maximum generation length may lead to different outcomes. In our experiments, as described in Section IV.A (Implementation), we deploy Llama-3.3 70B locally using Ollama with its default configuration and set the temperature to 0.0 to obtain more deterministic outputs, keeping all other parameters fixed across all experimental settings to mitigate this threat.

Randomness may also affect the effectiveness of \tech{}. However, using three repetitions in our experiments is sufficient to account for the stochasticity of LLM outputs. Empirical observations from RQ1 support this: on HumanEval-Java, \tech{} achieves average mutation scores of 89.9\%, 90.2\%, and 88.5\% across three runs (a range of 1.7\%), while on Leetcode-Java, the scores are 88.2\%, 88.0\%, and 91.1\% (a range of 2.1\%). Such differences do not affect our conclusions. 
For the baseline generation tool, \evo{}, we observed very similar mutation scores across three runs and all subjects in both datasets, with almost no variation. Since method-level subjects have a limited search space (around 40 lines of code per subject, as reported in Section IV.A), the evolutionary search guided by coverage converges quickly toward similar solutions across runs. 
Observations for \evovariant{}, a variant of \evo{} that maximizes mutation score, are consistent with those of \evo{}, for the same reasons.
Overall, these small differences indicate stable results across runs. 
Therefore, performing three repetitions is sufficient to obtain reproducible results. Reporting the average over three runs is also consistent with prior work on evaluating LLM-based approaches~\cite{yuan2024evaluating,falcao2025evaluating,gao2025trae}, including test generation. This description has been clarified in Section IV.C.

The threat to external validity mainly lies in the subjects and the target approaches. For subject selection, we reused benchmarks from existing publications and further enhanced subject diversity by evaluating our approach on 100 additional subjects. These were randomly sampled from a corpus of LeetCode algorithm solutions. In the future, we will evaluate \tech{} on more complicated subjects, e.g., Defects4J. Notably, Defects4J contains real-world defects, which can provide a stronger assessment of the practical effectiveness of our approach than synthetic defects simulated via mutation testing in this study.

Regarding the baseline approaches, we selected two representative methods: (1) \evo{}, a widely-used state-of-the-art search-based testing tool, and (2) \vanilla{}, a vanilla LLM-based prompt approach. These baselines allow us to evaluate whether incorporating mutation feedback in \tech{} significantly improves fault detection capability, as measured by mutation score.
In addition to the above compared methods, we also compared \tech{} with a variant of \evo{}. Following published guidelines~\cite{fraser2015achieving}, we enabled the fitness function for strong mutation testing in \evo{}. We conducted experiments on the two datasets using only the strong mutation fitness function. The average mutation scores for both datasets were 67.4\% and 58.1\%, respectively, which are similar to the results reported in our paper using the \evo{} default setting. 
Although we use Llama-3.3 as the base LLM in this study, our approach is not tied to any specific model. \tech{} is model-agnostic by design, and evaluating its performance relative to other LLMs is left for future work.

Several factors may affect the construct validity of our evaluation.


\noindent\textit{Equivalent Mutants.}
Our approach uses the \texttt{PITest} mutation testing framework, which includes mechanisms to reduce the creation of equivalent mutants. While this mitigates the problem to some extent, the remaining equivalent mutants can reduce \tech{}'s efficiency and prevent it from achieving a 100\% mutation score. \addminor{Since equivalent mutants affect all techniques equally, they do not affect our comparisons; addressing this well-known issue requires further investigation, which we leave for future work.}

\noindent\textit{Oracle Problem.}
We assume the original code is correct and only address assertion errors that cause Maven build failures. This simplification can lead to incorrect test oracles and false positives. 

\noindent\textit{Data Leakage.}
While some public code in benchmark datasets may appear in the LLM's training data (e.g., Llama-3.3), the specific mutants we generate are unlikely to be memorized by the LLM. Moreover, we compare \tech{} with a vanilla prompting baseline, and the observed performance gains suggest that our improvements stem from prompt engineering and strategy design rather than memorization.

\addminor{Another potential threat to construct validity stems from the unknown training data of the LLMs. Since the information about their training corpora is not publicly available, we cannot exclude the possibility that certain mutation patterns or code structures have been seen during pre-training. This may introduce biases, where the models perform worse on specific mutation operators that are less represented in their training data. To mitigate this concern, we use widely adopted, general-purpose LLMs rather than models specifically trained for mutation testing or program repair. Nevertheless, we acknowledge that the influence of training data can neither be confirmed nor ruled out and leave the investigation of this factor to future work.}

\section{Related Work} \label{sec:relate}

\textbf{Mutation-Guided Test Generation.} Recent studies have questioned the sufficiency of coverage as an evaluation criterion. As shown in~\cite{chekam2017empirical,andrews2006using,inozemtseva2014coverage}, and pointed in recent work from Meta~\cite{alshahwan2024automated,foster2025mutation}, high coverage does not necessarily correlate with strong defect detection capability. Mutation testing offers a more rigorous alternative by assessing how well tests detect injected faults (mutants)~\cite{chekam2017empirical}. Although some prior work~\cite{dakhel2024effective,foster2025mutation} has attempted to improve mutation scores during test generation, it does not fully explore the underlying limitations of LLMs in detecting and killing live mutants.
Our work aims to address this gap by incorporating mutation feedback into the prompt design process for LLMs, thereby enhancing the effectiveness of the generated test cases. In addition, we introduce an iterative generation process to push the limits of LLMs in killing mutants, providing insights into their current capabilities and boundaries. This exploration also highlights opportunities for future work to guide LLMs in generating test cases with stronger defect-detection capabilities.

\textbf{Other techniques.} 
Automatic unit test generation has long been a central research topic in software engineering. Traditional efforts have primarily focused on search-based and random testing techniques, with tools such as \evo{}~\cite{fraser2011evosuite} and Randoop~\cite{pacheco2007randoop} significantly reducing the developers' manual testing burden.
While \evo{} was originally designed to maximize code coverage, prior work~\cite{fraser2015achieving} has investigated enhancing mutation scores by modifying its fitness functions, an objective closely related to ours. We conducted experiments along these lines and discuss the results in Section IV.B. In short, results are not significantly different from those of the default setting of \evo{}. 

With the advent of large language models (LLMs), a new paradigm has emerged in test generation. 
Numerous recent surveys have summarized the rapid advancements in this area~\cite{fan2023large,wang2024software,hou2024large}; however, due to the large volume of related work, we only highlight a selection of representative examples in this section.
Early research explored fine-tuning LLMs for code-related tasks. As LLMs became more powerful and general-purpose, however, prompt-based approaches gained traction, enabling test generation without task-specific training. Several studies~\cite{siddiq2024using,yang2024evaluation,yuan2024evaluating,zhang2024testbench,li2023nuances,pizzorno2024coverup} have investigated prompt engineering strategies and mechanisms to improve test case generation effectiveness using commercial LLMs, notably OpenAI’s GPT series~\cite{achiam2023gpt,ouyang2022training,brown2020language}.

Given the high cost of commercial LLMs, and the increasing availability of high-quality open-source alternatives such as the Llama and DeepSeek series~\cite{grattafiori2024llama,liu2024deepseek}, recent research has shifted toward optimizing prompt-based effectiveness for open-source models~\cite{ouedraogo2024large,wang2025projecttest,deljouyi2024leveraging,wang2025towards,huang2023agentcoder,kim2025llamaresttest}. These models are becoming popular due to their accessibility and effectiveness in code-related tasks.

LLM-based test generation has been applied across a wide range of testing scenarios. Some studies focus on multilingual test generation across languages such as Python~\cite{wang2025projecttest,liu2024llm,xu2025clover,tian2025fixing}, Java~\cite{zhang2024testbench,wang2025projecttest,chen2024chatunitest,gu2024improving}, and others~\cite{wang2025projecttest,liu2024llm,zhang2025citywalk,cheng2025rug,wang2025fine}. Other works target specific testing goals, such as detecting particular bugs~\cite{liu2024llm,foster2025mutation}, API testing~\cite{cao2024automated,deng2025lrasgen,li2025llm,kim2025llamaresttest}, or generating test oracles~\cite{khandaker2025augmentest,dinella2022toga,binta2024togll}. LLMs have also been integrated with traditional testing techniques, including mocking~\cite{nan2025test,zhu2025understanding}, evolutionary algorithms~\cite{taherkhani2024epic}, and program analysis~\cite{gu2025llm,pan2025aster,ryan2024code,pan2024multi}. Evaluation in these works typically emphasizes code coverage metrics, such as branch and line coverage.

Unlike prior work, our approach focuses on automatically improving the mutation score with LLMs, based on mutation testing feedback, since such a score better reflects the fault-detection capability of test cases than code coverage~\cite{chekam2017empirical,foster2025mutation}. As discussed in Section~\ref{sec:example_gen}, high code coverage does not necessarily imply strong fault detection~\cite{chekam2017empirical}. However, as demonstrated in our experiments (Section~\ref{sec:eval_analysis}), improving mutation score often leads to increased code coverage as a by-product.

\section{Discussion} \label{sec:discuss}

We briefly discuss the limitations of our current work and outline directions for future research.

As described in Section~\ref{sec:eval} and \ref{sec:relate}, our experiments were conducted on independent, method-level subjects. This setup demonstrated how well LLMs perform at killing mutants when not confounded by broader program structures or dependencies. However, when transitioning to more complex datasets such as Defects4J, additional challenges arise, particularly in resolving dependencies to generate executable and meaningful test suites. We leave the investigation of such datasets for future work.

In addition, while we use Llama-3.3 as the base model in this study, \tech{} is inherently model-agnostic and does not rely on any specific LLM. Although our results demonstrate the effectiveness of the approach, other models may yield even better results. A more comprehensive evaluation across different LLMs is also left as future work.

Finally, although \tech{} achieves high mutation scores on the selected subjects, there remain open challenges in further improving these scores. For example, guiding LLMs to effectively kill mutants produced by certain mutation operators, such as \texttt{VoidMethodCalls}, remains difficult. In our evaluation on HumanEval-Java, this operator achieved only a 69.2\% kill ratio. Developing strategies to better handle such challenging mutants is an important direction for future research. We also observe that some generated test cases that are not repaired by the fixing stage are close to killing hard-to-detect mutants, but fail due to issues such as incorrect assertions or incomplete oracles. 
\add{As analyzed in Section III.B(c), 74\% of the unfixed test cases are caused by incorrect or incomplete oracles, while the remaining 26\% fail due to runtime exceptions (e.g., invalid inputs). This indicates that these unfixed tests warrant more complex oracle repair strategies.}
In addition, an important avenue for future work is to more efficiently characterize the upper bound of LLM mutation scores across diverse target programs. In particular, identifying how quickly an LLM approaches its performance limit and how this limit varies across program sizes, mutation operators, and code structures remains an open problem.

\section{Conclusion}
In this paper, we propose \tech{}, a mutation-guided, LLM-based approach for automatically generating unit tests with high fault detection capability. The core innovation of \tech{} lies in incorporating mutation feedback into the prompt to help LLMs generate tests that kill more mutants. In addition, two components—code summarization and code fixing—further enhance the approach's effectiveness.

We evaluated \tech{} on 104 subjects from the widely-used HumanEval-Java dataset and 100 subjects from Leetcode-Java, which we constructed from LeetCode algorithmic problem solutions. Experimental results show that \tech{} significantly outperforms both \evo{} and Llama-3.3 using a vanilla prompting strategy, achieving mutation scores of 89.5\% and 89.1\% on HumanEval-Java and Leetcode-Java, respectively. Since \tech{} aims to maximize the mutation score of generated test cases, we also compare it with a variant of \evo{} (\evovariant{}), where the fitness function of maximizing coverage is replaced with one that maximizes mutation score. The results show that \evovariant{} performs similarly to \evo{}, but both are outperformed by \tech{}.

Moreover, \tech{} employs an iterative generation process to push the limits of LLMs in killing mutants. \addminor{While \tech{} may produce some failing test cases, it is able to successfully repair around 50\% of them, which is considered a strong result in the recent software repair literature. We further analyze the causes of remaining failures, as well as the characteristics of live and uncovered mutants and the impact of different mutation operators, offering insights and directions for future research.}

\section{Acknowledgements}

We sincerely thank the anonymous reviewers for their valuable comments and suggestions, which have greatly helped improve the quality of this paper.

This publication has emanated from research jointly funded by Taighde Éireann -- Research Ireland under Grant No.~13/RC/2094\_2 and by Huawei Technologies Co., Ltd. Lionel Briand is also supported by the Natural Sciences and Engineering Research Council of Canada.

For the purpose of Open Access, the authors have applied a CC BY public copyright licence to any Author Accepted Manuscript version arising from this submission.

\bibliographystyle{IEEEtran}
\bibliography{ref}

@article{yuan2024evaluating,
  title={Evaluating and improving chatgpt for unit test generation},
  author={Yuan, Zhiqiang and Liu, Mingwei and Ding, Shiji and Wang, Kaixin and Chen, Yixuan and Peng, Xin and Lou, Yiling},
  journal={Proceedings of the ACM on Software Engineering},
  volume={1},
  number={FSE},
  pages={1703--1726},
  year={2024},
  publisher={ACM New York, NY, USA}
}

@article{alshahwan2024automated,
  title={Automated Unit Test Improvement using Large Language Models at Meta. CoRR abs/2402.09171 (2024)},
  author={Alshahwan, Nadia and Chheda, Jubin and Finegenova, Anastasia and Gokkaya, Beliz and Harman, Mark and Harper, Inna and Marginean, Alexandru and Sengupta, Shubho and Wang, Eddy},
  journal={arXiv preprint arXiv:2402.09171},
  volume={10},
  year={2024}
}

@inproceedings{chekam2017empirical,
  title={An empirical study on mutation, statement and branch coverage fault revelation that avoids the unreliable clean program assumption},
  author={Chekam, Thierry Titcheu and Papadakis, Mike and Le Traon, Yves and Harman, Mark},
  booktitle={2017 IEEE/ACM 39th International Conference on Software Engineering (ICSE)},
  pages={597--608},
  year={2017},
  organization={IEEE}
}

@article{foster2025mutation,
  title={Mutation-Guided LLM-based Test Generation at Meta},
  author={Foster, Christopher and Gulati, Abhishek and Harman, Mark and Harper, Inna and Mao, Ke and Ritchey, Jillian and Robert, Herv{\'e} and Sengupta, Shubho},
  journal={arXiv preprint arXiv:2501.12862},
  year={2025}
}

@article{gao2025prompt,
  title={The Prompt Alchemist: Automated LLM-Tailored Prompt Optimization for Test Case Generation},
  author={Gao, Shuzheng and Wang, Chaozheng and Gao, Cuiyun and Jiao, Xiaoqian and Chong, Chun Yong and Gao, Shan and Lyu, Michael},
  journal={arXiv preprint arXiv:2501.01329},
  year={2025}
}

@inproceedings{siddiq2024using,
  title={Using large language models to generate junit tests: An empirical study},
  author={Siddiq, Mohammed Latif and Da Silva Santos, Joanna Cecilia and Tanvir, Ridwanul Hasan and Ulfat, Noshin and Al Rifat, Fahmid and Carvalho Lopes, Vin{\'\i}cius},
  booktitle={Proceedings of the 28th International Conference on Evaluation and Assessment in Software Engineering},
  pages={313--322},
  year={2024}
}

@article{athiwaratkun2022multi,
  title={Multi-lingual evaluation of code generation models},
  author={Athiwaratkun, Ben and Gouda, Sanjay Krishna and Wang, Zijian and Li, Xiaopeng and Tian, Yuchen and Tan, Ming and Ahmad, Wasi Uddin and Wang, Shiqi and Sun, Qing and Shang, Mingyue and others},
  journal={arXiv preprint arXiv:2210.14868},
  year={2022}
}

@article{islam2405mapcoder,
  title={Mapcoder: Multi-agent code generation for competitive problem solving, 2024},
  author={Islam, Md Ashraful and Ali, Mohammed Eunus and Parvez, Md Rizwan},
  journal={URL https://arxiv. org/abs/2405},
  volume={11403}
}

@misc{pitest,
title = "PITest",
year = "Accessed: 2025",
url = {https://pitest.org}
}

@misc{jacoco,
title = "JaCoCo",
year = "Accessed: 2025",
url = {https://www.jacoco.org}
}

@misc{mutahunter,
title = "Mutahunter",
year = "Accessed: 2025",
url = {https://github.com/codeintegrity-ai/mutahunter}
}

@misc{ollama,
title = "Ollama",
year = "Accessed: 2025",
url = {https://ollama.com}
}

@misc{leetcode,
title = "LeetCode",
year = "Accessed: 2025",
url = {https://leetcode.com}
}

@misc{mutops,
title = "Description on mutation operators",
year = "Accessed: 2025",
url = {https://pitest.org/quickstart/mutators}
}

@article{grattafiori2024llama,
  title={The llama 3 herd of models},
  author={Grattafiori, Aaron and Dubey, Abhimanyu and Jauhri, Abhinav and Pandey, Abhinav and Kadian, Abhishek and Al-Dahle, Ahmad and Letman, Aiesha and Mathur, Akhil and Schelten, Alan and Vaughan, Alex and others},
  journal={arXiv preprint arXiv:2407.21783},
  year={2024}
}

@article{cao2024automated,
  title={Automated Test-Case Generation for REST APIs Using Model Inference Search Heuristic},
  author={Cao, Clinton and Panichella, Annibale and Verwer, Sicco},
  journal={arXiv preprint arXiv:2412.03420},
  year={2024}
}

@article{dakhel2024effective,
  title={Effective test generation using pre-trained large language models and mutation testing},
  author={Dakhel, Arghavan Moradi and Nikanjam, Amin and Majdinasab, Vahid and Khomh, Foutse and Desmarais, Michel C},
  journal={Information and Software Technology},
  volume={171},
  pages={107468},
  year={2024},
  publisher={Elsevier}
}

@inproceedings{fraser2011evosuite,
  title={Evosuite: automatic test suite generation for object-oriented software},
  author={Fraser, Gordon and Arcuri, Andrea},
  booktitle={Proceedings of the 19th ACM SIGSOFT symposium and the 13th European conference on Foundations of software engineering},
  pages={416--419},
  year={2011}
}

@inproceedings{pacheco2007randoop,
  title={Randoop: feedback-directed random testing for Java},
  author={Pacheco, Carlos and Ernst, Michael D},
  booktitle={Companion to the 22nd ACM SIGPLAN conference on Object-oriented programming systems and applications companion},
  pages={815--816},
  year={2007}
}

@article{vargha2000critique,
  title={A critique and improvement of the CL common language effect size statistics of McGraw and Wong},
  author={Vargha, Andr{\'a}s and Delaney, Harold D},
  journal={Journal of Educational and Behavioral Statistics},
  volume={25},
  number={2},
  pages={101--132},
  year={2000},
  publisher={Sage Publications Sage CA: Los Angeles, CA}
}

@inproceedings{arcuri2011practical,
  title={A practical guide for using statistical tests to assess randomized algorithms in software engineering},
  author={Arcuri, Andrea and Briand, Lionel},
  booktitle={Proceedings of the 33rd international conference on software engineering},
  pages={1--10},
  year={2011}
}

@article{achiam2023gpt,
  title={Gpt-4 technical report},
  author={Achiam, Josh and Adler, Steven and Agarwal, Sandhini and Ahmad, Lama and Akkaya, Ilge and Aleman, Florencia Leoni and Almeida, Diogo and Altenschmidt, Janko and Altman, Sam and Anadkat, Shyamal and others},
  journal={arXiv preprint arXiv:2303.08774},
  year={2023}
}

@article{brown2020language,
  title={Language models are few-shot learners},
  author={Brown, Tom and Mann, Benjamin and Ryder, Nick and Subbiah, Melanie and Kaplan, Jared D and Dhariwal, Prafulla and Neelakantan, Arvind and Shyam, Pranav and Sastry, Girish and Askell, Amanda and others},
  journal={Advances in neural information processing systems},
  volume={33},
  pages={1877--1901},
  year={2020}
}

@article{ouyang2022training,
  title={Training language models to follow instructions with human feedback},
  author={Ouyang, Long and Wu, Jeffrey and Jiang, Xu and Almeida, Diogo and Wainwright, Carroll and Mishkin, Pamela and Zhang, Chong and Agarwal, Sandhini and Slama, Katarina and Ray, Alex and others},
  journal={Advances in neural information processing systems},
  volume={35},
  pages={27730--27744},
  year={2022}
}

@article{deng2025lrasgen,
  title={LRASGen: LLM-based RESTful API Specification Generation},
  author={Deng, Sida and Huang, Rubing and Zhang, Man and Cui, Chenhui and Towey, Dave and Wang, Rongcun},
  journal={arXiv preprint arXiv:2504.16833},
  year={2025}
}

@inproceedings{yang2024evaluation,
  title={On the evaluation of large language models in unit test generation},
  author={Yang, Lin and Yang, Chen and Gao, Shutao and Wang, Weijing and Wang, Bo and Zhu, Qihao and Chu, Xiao and Zhou, Jianyi and Liang, Guangtai and Wang, Qianxiang and others},
  booktitle={Proceedings of the 39th IEEE/ACM International Conference on Automated Software Engineering},
  pages={1607--1619},
  year={2024}
}

@inproceedings{nan2025test,
  title={Test Intention Guided LLM-based Unit Test Generation},
  author={Nan, Zifan and Guo, Zhaoqiang and Liu, Kui and Xia, Xin},
  booktitle={2025 IEEE/ACM 47th International Conference on Software Engineering (ICSE)},
  pages={779--779},
  year={2025},
  organization={IEEE Computer Society}
}

@article{zhang2024testbench,
  title={TestBench: Evaluating Class-Level Test Case Generation Capability of Large Language Models},
  author={Zhang, Quanjun and Shang, Ye and Fang, Chunrong and Gu, Siqi and Zhou, Jianyi and Chen, Zhenyu},
  journal={arXiv preprint arXiv:2409.17561},
  year={2024}
}

@article{wang2025projecttest,
  title={ProjectTest: A Project-level Unit Test Generation Benchmark and Impact of Error Fixing Mechanisms},
  author={Wang, Yibo and Xia, Congying and Zhao, Wenting and Du, Jiangshu and Miao, Chunyu and Deng, Zhongfen and Yu, Philip S and Xing, Chen},
  journal={arXiv preprint arXiv:2502.06556},
  year={2025}
}

@article{liu2024llm,
  title={Llm-powered test case generation for detecting tricky bugs},
  author={Liu, Kaibo and Liu, Yiyang and Chen, Zhenpeng and Zhang, Jie M and Han, Yudong and Ma, Yun and Li, Ge and Huang, Gang},
  journal={arXiv preprint arXiv:2404.10304},
  year={2024}
}

@article{zhang2025citywalk,
  title={CITYWALK: Enhancing LLM-Based C++ Unit Test Generation via Project-Dependency Awareness and Language-Specific Knowledge},
  author={Zhang, Yuwei and Lu, Qingyuan and Liu, Kai and Dou, Wensheng and Zhu, Jiaxin and Qian, Li and Zhang, Chunxi and Lin, Zheng and Wei, Jun},
  journal={arXiv preprint arXiv:2501.16155},
  year={2025}
}

@article{gu2025llm,
  title={LLM Test Generation via Iterative Hybrid Program Analysis},
  author={Gu, Sijia and Nashid, Noor and Mesbah, Ali},
  journal={arXiv preprint arXiv:2503.13580},
  year={2025}
}

@article{pan2025aster,
  title={ASTER: Natural and Multi-language Unit Test Generation with LLMs},
  author={Pan, Rangeet and Kim, Myeongsoo and Krishna, Rahul and Pavuluri, Raju and Sinha, Saurabh},
  journal={arXiv preprint arXiv:2409.03093},
  year={2025}
}

@article{ryan2024code,
  title={Code-aware prompting: A study of coverage-guided test generation in regression setting using llm},
  author={Ryan, Gabriel and Jain, Siddhartha and Shang, Mingyue and Wang, Shiqi and Ma, Xiaofei and Ramanathan, Murali Krishna and Ray, Baishakhi},
  journal={Proceedings of the ACM on Software Engineering},
  volume={1},
  number={FSE},
  pages={951--971},
  year={2024},
  publisher={ACM New York, NY, USA}
}

@article{andrews2006using,
  title={Using mutation analysis for assessing and comparing testing coverage criteria},
  author={Andrews, James H and Briand, Lionel C and Labiche, Yvan and Namin, Akbar Siami},
  journal={IEEE Transactions on Software Engineering},
  volume={32},
  number={8},
  pages={608--624},
  year={2006},
  publisher={IEEE}
}

@article{xu2025clover,
  title={CLOVER: A Test Case Generation Benchmark with Coverage, Long-Context, and Verification},
  author={Xu, Jiacheng and Pang, Bo and Qu, Jin and Hayashi, Hiroaki and Xiong, Caiming and Zhou, Yingbo},
  journal={arXiv preprint arXiv:2502.08806},
  year={2025}
}

@article{khandaker2025augmentest,
  title={AugmenTest: Enhancing Tests with LLM-Driven Oracles},
  author={Khandaker, Shaker Mahmud and Kifetew, Fitsum and Prandi, Davide and Susi, Angelo},
  journal={arXiv preprint arXiv:2501.17461},
  year={2025}
}

@inproceedings{chen2024chatunitest,
  title={Chatunitest: A framework for llm-based test generation},
  author={Chen, Yinghao and Hu, Zehao and Zhi, Chen and Han, Junxiao and Deng, Shuiguang and Yin, Jianwei},
  booktitle={Companion Proceedings of the 32nd ACM International Conference on the Foundations of Software Engineering},
  pages={572--576},
  year={2024}
}

@article{gu2024improving,
  title={Improving LLM-based Unit test generation via Template-based Repair},
  author={Gu, Siqi and Fang, Chunrong and Zhang, Quanjun and Tian, Fangyuan and Zhou, Jianyi and Chen, Zhenyu},
  journal={arXiv preprint arXiv:2408.03095},
  year={2024}
}

@article{deljouyi2024leveraging,
  title={Leveraging large language models for enhancing the understandability of generated unit tests},
  author={Deljouyi, Amirhossein and Koohestani, Roham and Izadi, Maliheh and Zaidman, Andy},
  journal={arXiv preprint arXiv:2408.11710},
  year={2024}
}

@inproceedings{cheng2025rug,
  title={RUG: Turbo LLM for Rust Unit Test Generation},
  author={Cheng, Xiang and Sang, Fan and Zhai, Yizhuo and Zhang, Xiaokuan and Kim, Taesoo},
  booktitle={2025 IEEE/ACM 47th International Conference on Software Engineering (ICSE)},
  pages={634--634},
  year={2025},
  organization={IEEE Computer Society}
}

@article{wang2025fine,
  title={Fine-grained Testing for Autonomous Driving Software: a Study on Autoware with LLM-driven Unit Testing},
  author={Wang, Wenhan and Xie, Xuan and Huang, Yuheng and Wang, Renzhi and Chen, An Ran and Ma, Lei},
  journal={arXiv preprint arXiv:2501.09866},
  year={2025}
}

@article{li2025llm,
  title={LLM-assisted Mutation for Whitebox API Testing},
  author={Li, Jia and Shen, Jiacheng and Su, Yuxin and Lyu, Michael R},
  journal={arXiv preprint arXiv:2504.05738},
  year={2025}
}

@article{taherkhani2024epic,
  title={Epic: Cost-effective search-based prompt engineering of llms for code generation},
  author={Taherkhani, Hamed and Sepindband, Melika and Pham, Hung Viet and Wang, Song and Hemmati, Hadi},
  journal={arXiv preprint arXiv:2408.11198},
  year={2024}
}

@inproceedings{inozemtseva2014coverage,
  title={Coverage is not strongly correlated with test suite effectiveness},
  author={Inozemtseva, Laura and Holmes, Reid},
  booktitle={Proceedings of the 36th international conference on software engineering},
  pages={435--445},
  year={2014}
}

@article{liu2024deepseek,
  title={Deepseek-v3 technical report},
  author={Liu, Aixin and Feng, Bei and Xue, Bing and Wang, Bingxuan and Wu, Bochao and Lu, Chengda and Zhao, Chenggang and Deng, Chengqi and Zhang, Chenyu and Ruan, Chong and others},
  journal={arXiv preprint arXiv:2412.19437},
  year={2024}
}

@article{wang2025towards,
  title={Towards understanding the characteristics of code generation errors made by large language models},
  author={Wang, Zhijie and Zhou, Zijie and Da Song, Yuheng Huang and Chen, Shengmai and Ma, Lei and Zhang, Tianyi},
  journal={Preprint},
  year={2025}
}

@article{huang2023agentcoder,
  title={Agentcoder: Multi-agent-based code generation with iterative testing and optimisation},
  author={Huang, Dong and Zhang, Jie M and Luck, Michael and Bu, Qingwen and Qing, Yuhao and Cui, Heming},
  journal={arXiv preprint arXiv:2312.13010},
  year={2023}
}

@article{kim2025llamaresttest,
  title={LlamaRestTest: Effective REST API Testing with Small Language Models},
  author={Kim, Myeongsoo and Sinha, Saurabh and Orso, Alessandro},
  journal={arXiv preprint arXiv:2501.08598},
  year={2025}
}

@inproceedings{li2023nuances,
  title={Nuances are the key: Unlocking chatgpt to find failure-inducing tests with differential prompting},
  author={Li, Tsz-On and Zong, Wenxi and Wang, Yibo and Tian, Haoye and Wang, Ying and Cheung, Shing-Chi and Kramer, Jeff},
  booktitle={2023 38th IEEE/ACM International Conference on Automated Software Engineering (ASE)},
  pages={14--26},
  year={2023},
  organization={IEEE}
}

@inproceedings{tian2025fixing,
  title={Fixing Large Language Models' Specification Misunderstanding for Better Code Generation},
  author={Tian, Zhao and Chen, Junjie and Zhang, Xiangyu},
  booktitle={2025 IEEE/ACM 47th International Conference on Software Engineering (ICSE)},
  pages={645--645},
  year={2025},
  organization={IEEE Computer Society}
}

@article{zhu2025understanding,
  title={Understanding and Characterizing Mock Assertions in Unit Tests},
  author={ZHU, HENGCHENG and TERRAGNI, VALERIO and WEI, LILI and CHEUNG, SHING-CHI and WU, JIARONG and LIU, YEPANG},
  year={2025}
}

@article{ouedraogo2024large,
  title={Large-scale, Independent and Comprehensive study of the power of LLMs for test case generation},
  author={Ou{\'e}draogo, Wendk{\^u}uni C and Kabor{\'e}, Kader and Tian, Haoye and Song, Yewei and Koyuncu, Anil and Klein, Jacques and Lo, David and Bissyand{\'e}, Tegawend{\'e} F},
  journal={arXiv preprint arXiv:2407.00225},
  year={2024}
}

@inproceedings{dinella2022toga,
  title={Toga: A neural method for test oracle generation},
  author={Dinella, Elizabeth and Ryan, Gabriel and Mytkowicz, Todd and Lahiri, Shuvendu K},
  booktitle={Proceedings of the 44th International Conference on Software Engineering},
  pages={2130--2141},
  year={2022}
}

@article{binta2024togll,
  title={TOGLL: Correct and Strong Test Oracle Generation with LLMs},
  author={Binta Hossain, Soneya and Dwyer, Matthew},
  journal={arXiv e-prints},
  pages={arXiv--2405},
  year={2024}
}

@article{pizzorno2024coverup,
  title={Coverup: Coverage-guided llm-based test generation},
  author={Pizzorno, Juan Altmayer and Berger, Emery D},
  journal={arXiv preprint arXiv:2403.16218},
  year={2024}
}

@article{hou2024large,
  title={Large language models for software engineering: A systematic literature review},
  author={Hou, Xinyi and Zhao, Yanjie and Liu, Yue and Yang, Zhou and Wang, Kailong and Li, Li and Luo, Xiapu and Lo, David and Grundy, John and Wang, Haoyu},
  journal={ACM Transactions on Software Engineering and Methodology},
  volume={33},
  number={8},
  pages={1--79},
  year={2024},
  publisher={ACM New York, NY}
}

@article{wang2024software,
  title={Software testing with large language models: Survey, landscape, and vision},
  author={Wang, Junjie and Huang, Yuchao and Chen, Chunyang and Liu, Zhe and Wang, Song and Wang, Qing},
  journal={IEEE Transactions on Software Engineering},
  year={2024},
  publisher={IEEE}
}

@inproceedings{fan2023large,
  title={Large language models for software engineering: Survey and open problems},
  author={Fan, Angela and Gokkaya, Beliz and Harman, Mark and Lyubarskiy, Mitya and Sengupta, Shubho and Yoo, Shin and Zhang, Jie M},
  booktitle={2023 IEEE/ACM International Conference on Software Engineering: Future of Software Engineering (ICSE-FoSE)},
  pages={31--53},
  year={2023},
  organization={IEEE}
}

@article{pan2024multi,
  title={Multi-language unit test generation using llms},
  author={Pan, Rangeet and Kim, Myeongsoo and Krishna, Rahul and Pavuluri, Raju and Sinha, Saurabh},
  journal={arXiv preprint arXiv:2409.03093},
  year={2024}
}

@article{rojas2016seeding,
  title={Seeding strategies in search-based unit test generation},
  author={Rojas, Jos{\'e} Miguel and Fraser, Gordon and Arcuri, Andrea},
  journal={Software Testing, Verification and Reliability},
  volume={26},
  number={5},
  pages={366--401},
  year={2016},
  publisher={Wiley Online Library}
}

@inproceedings{rojas2015combining,
  title={Combining multiple coverage criteria in search-based unit test generation},
  author={Rojas, Jos{\'e} Miguel and Campos, Jos{\'e} and Vivanti, Mattia and Fraser, Gordon and Arcuri, Andrea},
  booktitle={Search-Based Software Engineering: 7th International Symposium, SSBSE 2015, Bergamo, Italy, September 5-7, 2015, Proceedings 7},
  pages={93--108},
  year={2015},
  organization={Springer}
}

@inproceedings{vivanti2013search,
  title={Search-based data-flow test generation},
  author={Vivanti, Mattia and Mis, Andre and Gorla, Alessandra and Fraser, Gordon},
  booktitle={2013 IEEE 24th International Symposium on Software Reliability Engineering (ISSRE)},
  pages={370--379},
  year={2013},
  organization={IEEE}
}

@inproceedings{pacheco2007feedback,
  title={Feedback-directed random test generation},
  author={Pacheco, Carlos and Lahiri, Shuvendu K and Ernst, Michael D and Ball, Thomas},
  booktitle={29th International Conference on Software Engineering (ICSE'07)},
  pages={75--84},
  year={2007},
  organization={IEEE}
}

@inproceedings{oriat2005jartege,
  title={Jartege: a tool for random generation of unit tests for java classes},
  author={Oriat, Catherine},
  booktitle={International Conference on the Quality of Software Architectures},
  pages={242--256},
  year={2005},
  organization={Springer}
}

@article{andrews2011genetic,
  title={Genetic algorithms for randomized unit testing},
  author={Andrews, James H and Menzies, Tim and Li, Felix CH},
  journal={IEEE Transactions on Software Engineering},
  volume={37},
  number={1},
  pages={80--94},
  year={2011},
  publisher={IEEE}
}

@article{hamlet1994random,
  title={Random testing},
  author={Hamlet, Richard},
  journal={Encyclopedia of software Engineering},
  volume={2},
  pages={971--978},
  year={1994},
  publisher={Citeseer}
}

@book{naik2011software,
  title={Software testing and quality assurance: theory and practice},
  author={Naik, Kshirasagar and Tripathy, Priyadarshi},
  year={2011},
  publisher={John Wiley \& Sons}
}

@article{fraser2014large,
  title={A large-scale evaluation of automated unit test generation using evosuite},
  author={Fraser, Gordon and Arcuri, Andrea},
  journal={ACM Transactions on Software Engineering and Methodology (TOSEM)},
  volume={24},
  number={2},
  pages={1--42},
  year={2014},
  publisher={ACM New York, NY, USA}
}

@article{corno2004automatic,
  title={Automatic test program generation: a case study},
  author={Corno, Fulvio and S{\'a}nchez, Ernesto and Reorda, Matteo Sonza and Squillero, Giovanni},
  journal={IEEE Design \& Test of Computers},
  volume={21},
  number={2},
  pages={102--109},
  year={2004},
  publisher={IEEE}
}

@article{anand2013orchestrated,
  title={An orchestrated survey of methodologies for automated software test case generation},
  author={Anand, Saswat and Burke, Edmund K and Chen, Tsong Yueh and Clark, John and Cohen, Myra B and Grieskamp, Wolfgang and Harman, Mark and Harrold, Mary Jean and McMinn, Phil and Bertolino, Antonia and others},
  journal={Journal of systems and software},
  volume={86},
  number={8},
  pages={1978--2001},
  year={2013},
  publisher={Elsevier}
}

@article{alagarsamy2024a3test,
  title={A3test: Assertion-augmented automated test case generation},
  author={Alagarsamy, Saranya and Tantithamthavorn, Chakkrit and Aleti, Aldeida},
  journal={Information and Software Technology},
  volume={176},
  pages={107565},
  year={2024},
  publisher={Elsevier}
}

@inproceedings{zheng2023codegeex,
  title={Codegeex: A pre-trained model for code generation with multilingual benchmarking on humaneval-x},
  author={Zheng, Qinkai and Xia, Xiao and Zou, Xu and Dong, Yuxiao and Wang, Shan and Xue, Yufei and Shen, Lei and Wang, Zihan and Wang, Andi and Li, Yang and others},
  booktitle={Proceedings of the 29th ACM SIGKDD Conference on Knowledge Discovery and Data Mining},
  pages={5673--5684},
  year={2023}
}

@article{chen2021evaluating,
  title={Evaluating large language models trained on code},
  author={Chen, Mark and Tworek, Jerry and Jun, Heewoo and Yuan, Qiming and Pinto, Henrique Ponde De Oliveira and Kaplan, Jared and Edwards, Harri and Burda, Yuri and Joseph, Nicholas and Brockman, Greg and others},
  journal={arXiv preprint arXiv:2107.03374},
  year={2021}
}

@article{fraser2015achieving,
  title={Achieving scalable mutation-based generation of whole test suites},
  author={Fraser, Gordon and Arcuri, Andrea},
  journal={Empirical Software Engineering},
  volume={20},
  number={3},
  pages={783--812},
  year={2015},
  publisher={Springer}
}

@inproceedings{just2014defects4j,
  title={Defects4J: A database of existing faults to enable controlled testing studies for Java programs},
  author={Just, Ren{\'e} and Jalali, Darioush and Ernst, Michael D},
  booktitle={Proceedings of the 2014 international symposium on software testing and analysis},
  pages={437--440},
  year={2014}
}

@article{wang2024comprehensive,
  title={A Comprehensive Study on Large Language Models for Mutation Testing},
  author={Wang, Bo and Chen, Mingda and Deng, Ming and Lin, Youfang and Harman, Mark and Papadakis, Mike and Zhang, Jie M},
  journal={arXiv preprint arXiv:2406.09843},
  year={2024}
}

@article{jia2010analysis,
  title={An analysis and survey of the development of mutation testing},
  author={Jia, Yue and Harman, Mark},
  journal={IEEE transactions on software engineering},
  volume={37},
  number={5},
  pages={649--678},
  year={2010},
  publisher={IEEE}
}

@article{falcao2025evaluating,
  title={Evaluating the effectiveness of LLM-based interoperability},
  author={Falc{\~a}o, Rodrigo and Schweitzer, Stefan and Siebert, Julien and Calvet, Emily and Elberzhager, Frank},
  journal={arXiv preprint arXiv:2510.23893},
  year={2025}
}

@article{gao2025trae,
  title={Trae agent: An llm-based agent for software engineering with test-time scaling},
  author={Gao, Pengfei and Tian, Zhao and Meng, Xiangxin and Wang, Xinchen and Hu, Ruida and Xiao, Yuanan and Liu, Yizhou and Zhang, Zhao and Chen, Junjie and Gao, Cuiyun and others},
  journal={arXiv preprint arXiv:2507.23370},
  year={2025}
}

@misc{cc,
      author = {TIOBE Organization},
      title = {Which programming language produces the most complex code?},
      howpublished = {\url{https://www.tiobe.com/knowledge/article/which-programming-language-produces-the-most-complex-code/}},
      note = {Accessed: 22 Nov, 2025},
      year = {2025}
    }

@article{ccpaper,
  title={Evaluating the Dependency Between Cyclomatic Complexity and Response For Class},
  author={Stavtsev, Maxim and Bugayenko, Yegor},
  journal={arXiv preprint arXiv:2410.06416},
  year={2024}
}

@article{yang2025survey,
  title={A survey of LLM-based automated program repair: Taxonomies, design paradigms, and applications},
  author={Yang, Boyang and Cai, Zijian and Liu, Fengling and Le, Bach and Zhang, Lingming and Bissyand{\'e}, Tegawend{\'e} F and Liu, Yang and Tian, Haoye},
  journal={arXiv preprint arXiv:2506.23749},
  year={2025}
}

@inproceedings{xia2024automated,
  title={Automated program repair via conversation: Fixing 162 out of 337 bugs for \$0.42 each using chatgpt},
  author={Xia, Chunqiu Steven and Zhang, Lingming},
  journal={Proceedings of the 33rd ACM SIGSOFT International Symposium on Software Testing and Analysis},
  pages={819--831},
  year={2024}
}

@article{gu2024testart,
  title={Testart: Improving llm-based unit testing via co-evolution of automated generation and repair iteration},
  author={Gu, Siqi and Zhang, Quanjun and Li, Kecheng and Fang, Chunrong and Tian, Fangyuan and Zhu, Liuchuan and Zhou, Jianyi and Chen, Zhenyu},
  journal={arXiv preprint arXiv:2408.03095},
  year={2024}
}

\newpage

\end{document}